\documentclass[]{tCPH2e}
\usepackage{color}

\begin{document}
\doi{0000}
 \issn{0000}
\issnp{0000}

\jvol{00} \jnum{00} \jyear{00} \jmonth{00}

\markboth{Nader Haghighipour}{Contemporary Physics}

\articletype{REVIEW}

\title{{\itshape Super-Earths: A New Class of Planetary Bodies}}

\author{Nader Haghighipour$^{a}$$^{\ast}$
\thanks{$^\ast$Corresponding author. Email: nader@ifa.hawaii.edu
\vspace{6pt}} \\
\vspace{6pt}  
$^{a}${\em{Institute for Astronomy and NASA Astrobiology Institute,\\ 
University of Hawaii, Honolulu, HI 96822, USA}}\\
\vspace{6pt}
\received{0000} }

\maketitle

\begin{abstract}

Super-Earths, a class of planetary bodies with masses ranging from a few
Earth-masses to slightly smaller than Uranus, 
have recently found a special place in the exoplanetary science. 
Being slightly larger than a typical terrestrial planet, super-Earths may have 
physical and dynamical characteristics similar to those of Earth 
whereas unlike terrestrial planets, they are relatively easier to detect. 
Because of their sizes, super-Earths can maintain moderate atmospheres and
possibly dynamic interiors with plate tectonics.
They also seem to be more
common around low-mass stars where the habitable zone is in closer distances.
This article presents a review of the current state of 
research on super-Earths, and discusses the models of the formation, dynamical 
evolution, and possible habitability of these objects. Given the recent advances in
detection techniques, the detectability of super-Earths is also discussed, 
and a review of the prospects of their detection in the habitable zones of 
low-mass stars is presented.
\bigskip

\bigskip
\bigskip

\end{abstract}

\section{Introduction}

It was almost 500 years ago when the Italian philosopher,
Giordano Bruno, discussed the possibility of the existence of planets around
other stars and presented the idea of ``countless suns and countless earths''
\footnote{From the quote by the Italian philosopher, Giordano
Bruno (1548-1600): ''There are countless suns and countless earths all 
rotating around 
their suns in exactly the same way as the seven planets of our system. 
We see only the suns because they are largest bodies and are luminous, but
their planets remain invisible to us because they are smaller and 
non-luminous. These countless worlds in the universe are no worse and no 
less inhabited than our Earth.'' Bruno was burned alive because of his 
ideas.}. 
Since then, as the science of astronomy progressed, it became more and more evident
that the Sun and our solar system are not unique,
and there must be many planets that revolve around other stars.
For centuries astronomers tried tirelessly to detect such {\it extrasolar}
planetary systems. However, until two decades ago, their efforts were rendered
fruitless--their detection techniques had not reached the level of sensitivity 
that was necessary to identify a planetary body either directly, or through 
its perturbation on its host star.

Thanks to advances in observation and detection technologies, in the 
past two decades this trend changed. Measurements of the shifts in the spectrum of 
the light of a star due to its radial velocity that is caused by the
gravitational attraction of a massive companion enabled astronomers to
identify many planetary bodies around nearby stars. The 
{\it Precision Radial Velocity Technique}, also known as {\it Doppler
Velocimetry} (Figures 1) has been successful in identifying now more than 
500 planets including the first exoplanetary body, a 4.7 Jupiter-mass object 
in a 4-day orbit around the Sun-like star 51 Pegasi \cite{51Peg}, 
and possibly an Earth-sized planet in the habitable zone 
of the near-by star Gliese 581 \cite{Vogt10}\footnote{It is important to note
that the first planets outside of our solar system were discovered  
around pulsar PSR B 1257+12 by Wolszczan \& Frail in 1992 \cite{Wolszczan92}.}. 

Technological advances also enabled astronomers to detect planetary bodies
by measuring the dimming of the light of a star due to a passing planetary companion.
This technique, known as {\it Transit Photometry}, has been successful in detecting now more 
than 130 planets. Figure 2 shows the schematics of this technique. As an example,
the actual light curve of the star HD 209458, the first star for which a planetary transit was
detected, is also shown. The transiting planet in this system is a
0.64 Jupiter-mass object in a 3.5-day orbit \cite{hd209458}.
In addition to the detection of planets, transit photometry 
has also enabled astronomers to determine the size, 
density \cite{Leger09,Charbonneau09},
and in some cases, the chemical elements in the atmospheres of transiting 
planets \cite{Tinetti07}. 

Other detection techniques such as microlensing, 
where the gravitational fields of a star 
and its planetary companion create magnifying effects 
of the light of a background source
(Figure 3) \cite {Gould06,Beaulieu06}, transit timing variations method, where
the gravitational perturbation of an object creates variations in the time and duration 
of the transits of a close-in planet \cite{Holman10,Lissauer11}, and 
direct imaging (Figures 4 and 5) \cite{Kalas08,Marois08,Marois10} have also been successful in 
detecting extrasolar planets. We refer the reader to the 
extrasolar planets encyclopedia at
{\tt http://exoplanet.eu}, and exoplanet data explorer at {\tt http://exoplanets.org/}
for more information.

\begin{figure}
\begin{center}
\begin{minipage}{150mm}
\center{
\resizebox*{11cm}{!}{\includegraphics{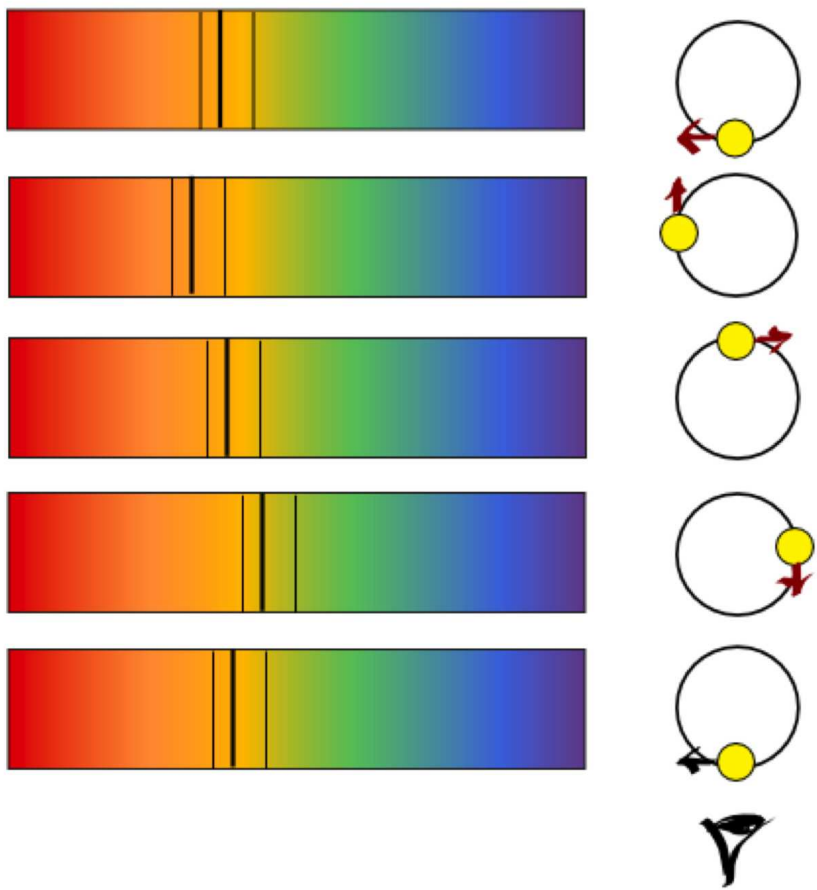}}}%
\vskip 5pt
\caption{Schematic illustration of Doppler shift of stellar light. As shown by the
second panel from top, the light is red-shifted when the star is moving
away from the observer. When the star moves towards the observer (second panel
from bottom), light is shifted towards blue. By measuring this Doppler shift, astronomers
can determine the semimajor axis, eccentricity, and minimum mass of the unseen planet.
Figure from spheroid.wordpress.com .}%
\label{RV2}
\end{minipage}
\end{center}
\end{figure}

\begin{figure}
\begin{center}
\begin{minipage}{150mm}
\hskip 50pt
\resizebox*{12cm}{!}{\includegraphics{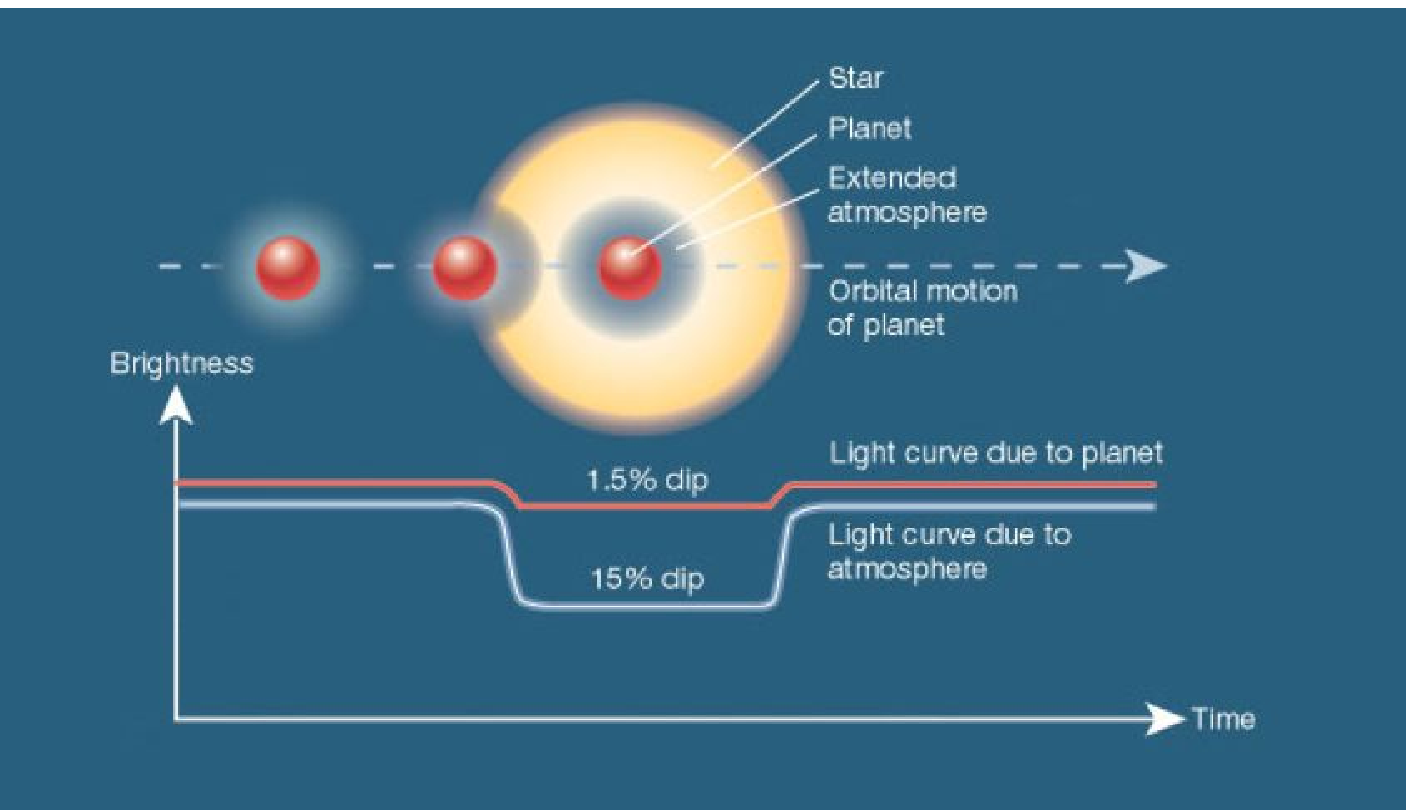}}
\vskip 5pt
\hskip 15pt
\resizebox*{13cm}{!}{\includegraphics{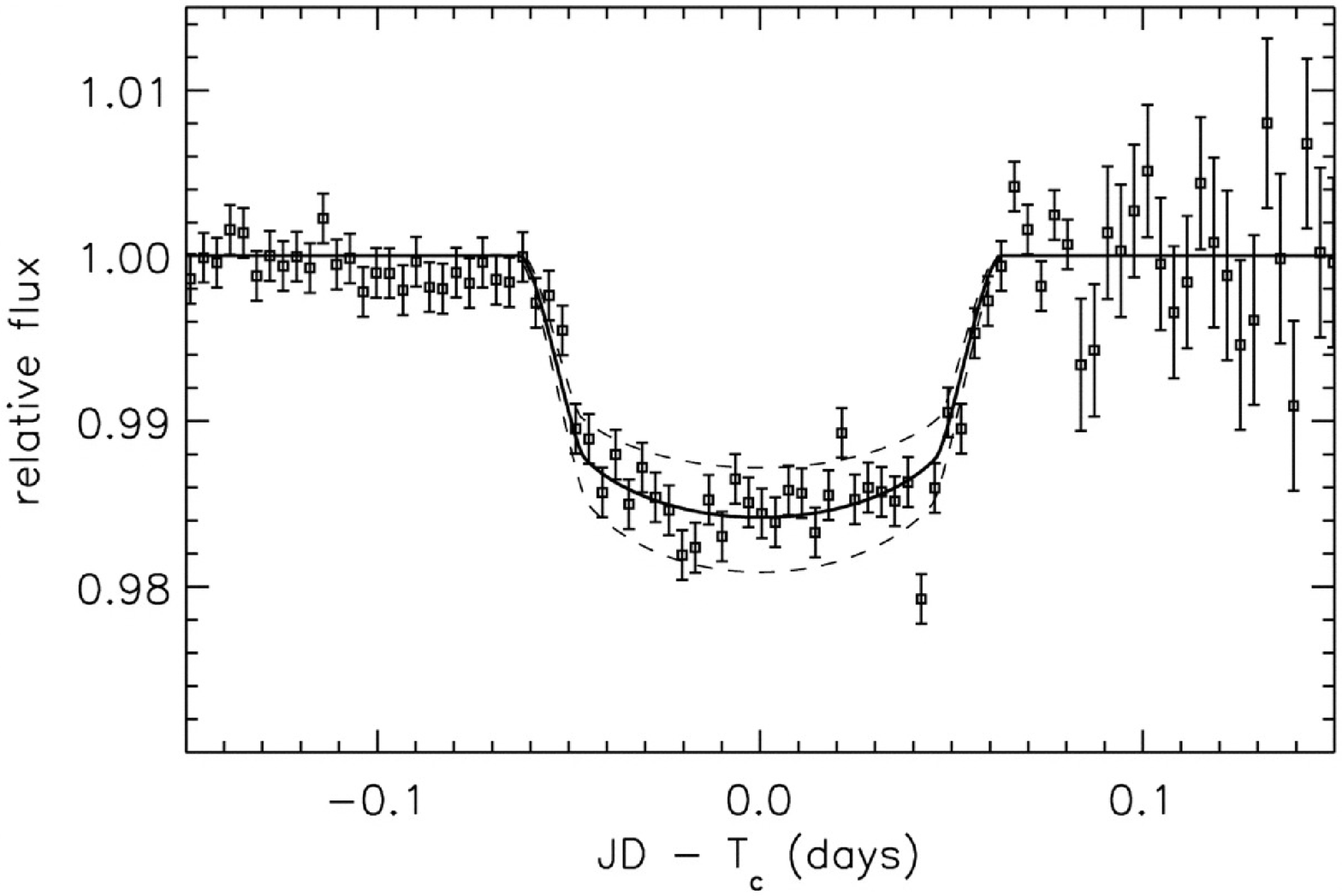}}%
\vskip 5pt
\caption{Top: Schematic illustration of a planetary transit. Bottom: Actual light curve
of HD 209458 caused by the transiting of its 0.64 Jupiter-mass planet \cite{hd209458}.
Figure from \cite{hd209458} with the permission of AAS.}%
\label{Transit}
\end{minipage}
\end{center}
\end{figure}

To-date the number of detected extrasolar planets exceeds
550. Almost all these planets depict physical and dynamical 
characteristics that are unlike those of the planets in our solar system. 
While in the solar system, giant planets such as Jupiter and Saturn are in 
large orbits, and smaller planets such as Earth and Venus are closer in,
many extrasolar planetary systems are host to Jupiter-like or larger
bodies in orbits smaller than the orbit of Mercury to the Sun. Also,
unlike in our solar system where planetary orbits are almost circular,
the orbits of many extrasolar planets are considerably elliptical.
These unexpected dynamical characteristics of exoplanets have had profound effects
on our views of the formation and dynamical evolution of planetary systems.
The theories of planet formation, which have been primarily developed 
to explain the formation of the planets of our solar system, are now constantly
revisited and their applicability to exoplanetary bodies are continuously challenged.

\begin{figure}
\begin{center}
\begin{minipage}{150mm}
\center{
\resizebox*{11cm}{!}{\includegraphics{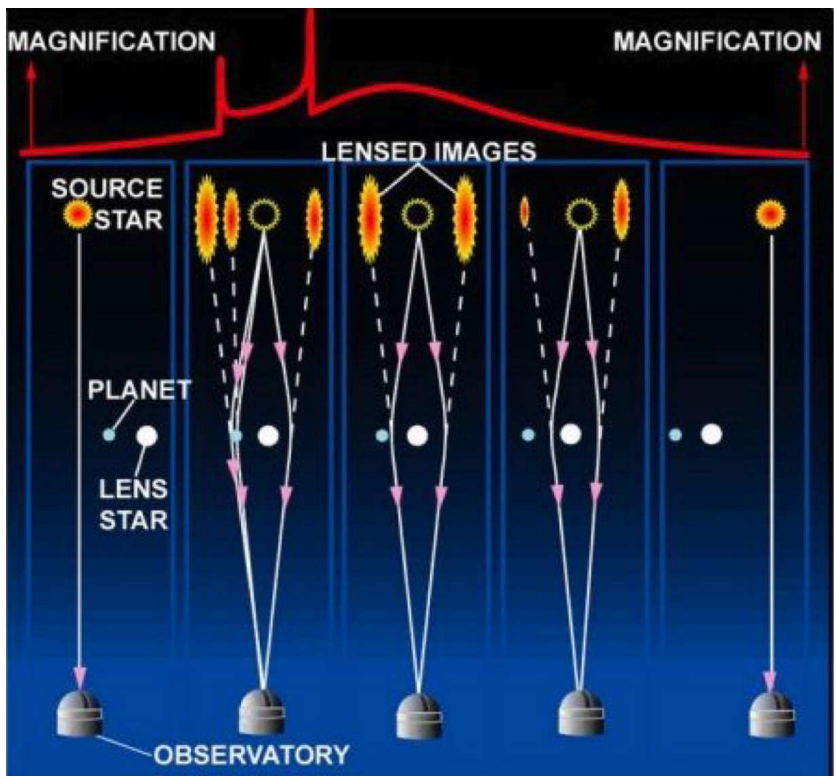}}}%
\vskip 5pt
\caption{Detection of a planet using gravitational microlensing. The time sequence 
can be taken from either left or right. The second panel from left shows
the lensing of a background source by a star and its planet. As the planet moves
away from the lensing state, the image of the background source is only
lensed by the planet-hosting star.
Figure courtesy of D. Bennett and Lockheed Martin Space Systems.}%
\label{Microlensing}
\end{minipage}
\end{center}
\end{figure}

The complexities of extrasolar planetary systems are not limited to 
their orbital dynamics. The physical characteristics of many of these 
objects are also different from the planets of our solar system. While in our
solar system, terrestrial and (gas- and ice-) giant planets form two distinct
classes of objects with two distinct ranges of masses
(giant planets are one to two orders of magnitude more massive than terrestrial
planets), several extrasolar planets have been discovered with 
masses in an intermediate range from a few Earth-masses to slightly smaller
than Uranus. Dubbed as {\it Super-Earths}, 
these objects form a new class of planetary bodies with physical and
dynamical characteristics that may be different from those of the
terrestrial planets and yet significant for habitability 
and planet formation theories.
This paper presents a review of the physical and dynamical 
characteristics of these objects. 

The first super-Earth was discovered by
Beaulieu et al. (2006; \cite{Beaulieu06}) using the microlensing technique.
To-date, the number of these objects has passed 30.
Table 1 shows the masses and orbital elements of these 
bodies\footnote{Note that Table 1 does not include the three terrestrial-class 
planets around the pulsar PSR 1257+12.}. Two of the more prominent super-Earths are 
CoRoT-7b, the 7th planet discovered by CoRoT (COnvection, ROtation and planetary 
Transits) space telescope with a mass of 2.3-8 Earth-masses 
\cite{Leger09,Queloz09,Hatzes10,Hatzes11},
and GJ 1214b, the first super-Earth discovered by transit photometry around an M star 
with a mass of 5.69 Earth-masses \cite{Charbonneau09}.
These two objects are the first super-Earths for which the values of mass and radius
have been measured (CoRoT-7b: 1.65 Earth-radii, GJ 1214b: 2.7 Earth-radii). 
This is a major achievement and a great milestone in the field of exoplanetary science
 which for the first time
allows for estimating the density of an extrasolar planet and developing models 
for its interior dynamics. 

The semimajor axes of the majority of super-Earths are smaller than 
0.2 AU and their eccentricities range from 0 to 0.4. This orbital 
diversity, combined with the values of the masses of these objects,
has made super-Earths a particularly important class of extrasolar planetary bodies.
The larger-than-terrestrial sizes and masses of super-Earths point to the less challenging
detection of these objects compared to the detection of Earth-sized planets.
They also suggest that super-Earths may have dynamic 
interiors and be able to develop 
and maintain moderate atmospheres--two conditions that would render super-Earths 
potentially habitable if their orbits are in the habitable zones of their host stars.

Although the close-in orbits of super-Earths pose a challenge to
the planet formation theories (many efforts have been made to explain the formation
of these objects in close-in orbits, and several models have been developed. 
However, this issue is still unresolved.), the physical characteristics of these objects, 
namely their densities,
when considered within the context of different formation scenarios, 
present a potential pathway for differentiating  between different planet formation models.
In that respect, the study of super-Earths plays an important role in identifying 
the most viable planet formation mechanism. 
In this paper, we discuss these issues and review the current state
of research on the formation, interior dynamics, and 
atmospheric evolution of super-Earths.
We also  review the prospects of the detection of these 
objects using ground- and space-based
telescopes as potential targets for searching for extrasolar habitable planets.

\begin{table}
\tbl{Currently known extrasolar planets with masses up to 10 Earth-masses. 
The quantities $M, P, a$ and $e$ represent the mass (in terms of Earth's mass $M_\oplus$), 
orbital period, semimajor axis, and orbital eccentricity of the planet. The mass 
of the central star is shown by $M_*$ and is given in the units of solar-masses $(M_\odot)$.}
{\begin{tabular}{@{}lllllll}
\toprule 
Planet & $M (M_\oplus$) & $P$ (day) & $a$ (AU)& 
$\>e$ & Stellar Type &  ${M_*} (M_\odot)$ \\
\colrule 
GL 581 e  &$\>\>$  1.70  & $\>\>$ 3.15  & $\>\>$ 0.03  &  0.0  & $\>\>$$\>\>$ M3  &$\>\>$  0.31 \\
Kepler-11f &$\>\>$ 2.30 &$\>\>$ 46.69 &$\>\>$ 0.25 & 0.0 & $\>\>$$\>\>$ G &$\>\>$ 0.95 \\
GL 581 g  &$\>\>$  3.10  &$\>\>$  36.56  &$\>\>$  0.15  &  0.0  &$\>\>$$\>\>$  M3  &$\>\>$  0.31\\
MOA-2007-BLG-192-L b  & $\>\>$ 3.18 &  &$\>\>$ 0.62  &  &  &$\>\>$ 0.06   \\
HD 156668 b  & $\>\>$ 4.16  & $\>\>$ 4.65  & $\>\>$ 0.05  &  0.0  &$\>\>$$\>\>$  K2  &  \\
HD 40307 b  &$\>\>$  4.19  & $\>\>$ 4.31  &$\>\>$  0.05  &  0.0  &$\>\>$$\>\>$  K2.5 V  &  \\
Kepler-11b &$\>\>$ 4.30 & $\>\>$ 10.3 & $\>\>$ 0.09 & 0.0 & $\>\>$$\>\>$ G & $\>\>$ 0.95 \\
Kepler-10b &$\>\>$ 4.56 & $\>\>$ 0.84 & $\>\>$ 0.02 &  0.0 & $\>\>$$\>\>$ G & $\>\>$ 0.89 \\
CoRoT-7b  &$\>\>$  4.80  & $\>\>$ 0.85  & $\>\>$ 0.02  &  0.0 & $\>\>$$\>\>$ K0 V  &$\>\>$  0.93 \\
61 Vir b  & $\>\>$ 5.08 & $\>\>$ 4.21  &$\>\>$  0.05  &  0.12  & $\>\>$$\>\>$ G5 V  &$\>\>$  0.95\\
GL 581 c  &$\>\>$  5.60  &$\>\>$  12.92  &$\>\>$  0.07  &  0.0  &$\>\>$$\>\>$  M3  &$\>\>$  0.31\\
HD 215497 b  &$\>\>$  5.40  &$\>\>$  3.93  &  &  & $\>\>$$\>\>$ K3 V  & $\>\>$ 0.87   \\
OGLE-05-390L b  &$\>\>$  5.40  &$\>\>$  3500  & $\>\>$ 2.10  &  &$\>\>$$\>\>$ M  &$\>\>$  0.22 \\
GJ 1214 b  &$\>\>$  5.69  &$\>\>$  1.58  &$\>\>$  0.01  &  0.27  &  &$\>\>$ 0.16   \\
GJ 667C b  & $\>\>$ 5.72  &$\>\>$  7.00  &  &  &$\>\>$$\>\>$ M1.5  &  \\
GJ 433 b  &$\>\>$  6.04  &$\>\>$  7.00  &  &  &$\>\>$$\>\>$ M1.5  &  \\
Kepler-11d &$\>\>$ 6.1 &$\>\>$ 22.68 &$\>\>$ 0.159 & 0 & $\>\>$$\>\>$ G & $\>\>$ 0.95\\
GJ 876 d  &$\>\>$  6.36  &$\>\>$  1.94  &$\>\>$  0.02  &  0.14  &$\>\>$$\>\>$  M4 V  &$\>\>$ 0.33\\
HD 40307 c  &$\>\>$  6.86  &$\>\>$  9.62  &  $\>\>$ 0.08  &  0.0  & $\>\>$$\>\>$ K2.5 V  &  \\
GL 581 d  &$\>\>$  5.60  &$\>\>$  66.87  &$\>\>$  0.22  &  0.0  &$\>\>$$\>\>$  M3  &$\>\>$  0.31\\
GL 581 f  &$\>\>$  7.00  &$\>\>$  433  &$\>\>$  0.76  &  0.0  &$\>\>$$\>\>$  M3  &$\>\>$  0.31\\
HD 181433 b  & $\>\>$ 7.56  &$\>\>$  9.37  &$\>\>$  0.08 & 0.39  &$\>\>$$\>\>$  K3I V &$\>\>$ 0.78\\
HD 1461 b  &$\>\>$  7.59  &$\>\>$  5.77  &$\>\>$  0.06  &  0.14  & $\>\>$$\>\>$ G0 V &$\>\>$ 1.08 \\
55 Cnc e  &$\>\>$ 7.63  &$\>\>$  2.82  &$\>\>$  0.04  &  0.07  & $\>\>$$\>\>$ G8 V  &$\>\>$  1.03 \\
CoRoT-7c  &$\>\>$  8.39  &$\>\>$  3.69  &$\>\>$  0.05  &  0.0  & $\>\>$$\>\>$ K0 V  &$\>\>$  0.93\\
Kepler-11e &$\>\>$ 8.40 &$\>\>$ 31.99 &$\>\>$ 0.19 & 0.0 & $\>\>$$\>\>$ G & $\>\>$0.95 \\ 
HD 285968 b  &$\>\>$ 8.42  &$\>\>$  8.78  &$\>\>$  0.07  &  0.0 &$\>\>$$\>\>$  M2.5 V & $\>\>$ 0.49\\
HD 40307 d  &$\>\>$  9.15 &$\>\>$  20.46  &$\>\>$  0.13  &  0.0  &$\>\>$$\>\>$  K2.5 V  &  \\
HD 7924 b  &$\>\>$  9.22  &$\>\>$  5.39  &$\>\>$  0.06  &  0.17  & $\>\>$$\>\>$ K0 V &$\>\>$ 0.83 \\
HD 69830 b  &$\>\>$  10.49  &$\>\>$  8.67  &$\>\>$  0.08  &  0.1  &$\>\>$$\>\>$ K0 V &$\>\>$ 0.86 \\
HD 160691 c  &$\>\>$ 10.56 &$\>\>$ 9.64 &$\>\>$ 0.09 & 0.17  &$\>\>$$\>\>$ G3 IV-V &$\>\>$ 1.08 \\
\botrule
\end{tabular}}
\end{table}

\begin{figure}
\begin{center}
\begin{minipage}{150mm}
\center{
\resizebox*{12cm}{!}{\includegraphics{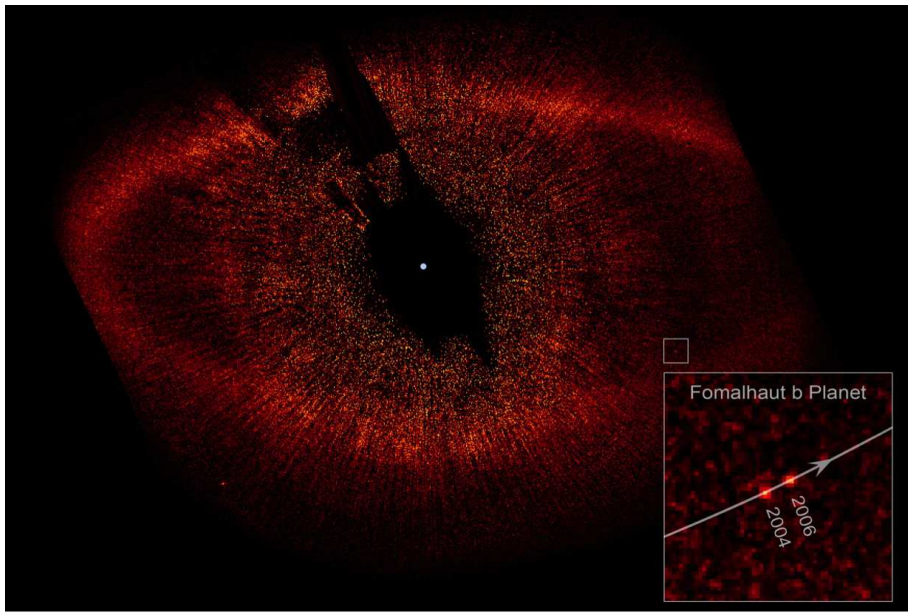}}}%
\caption{Image of the giant planet of Fomalhault by Kalas et al
(2008; \cite{Kalas08}). The central star is a 2 solar-masses A star at 25 light-years
from the Sun. The planet has a mass of  approximately 4 Jupiter-masses.
It orbits the central star at 115 AU with an eccentricity of 0.11. Figure
from \cite{Kalas08} with the permission of AAAS.}%
\label{Imaging1}
\end{minipage}
\end{center}
\end{figure}

\begin{figure}
\begin{center}
\begin{minipage}{150mm}
\center{
\resizebox*{12cm}{!}{\includegraphics{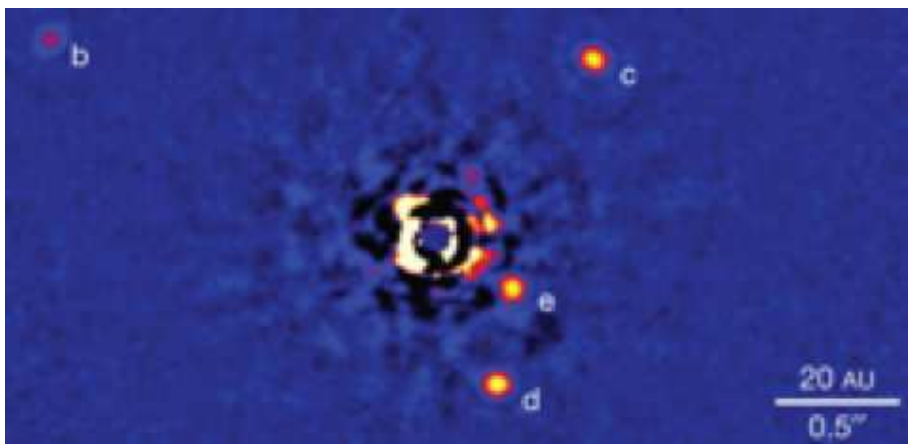}}}%
\vskip 5pt
\caption{Planetary system of HR 8799 imaged by Marois et al
(2010; \cite{Marois10}). The central star is of spectral type A
with a mass of 1.5 solar-masses at a distance of 128 light-years
from the Sun. The planets have the masses of $M_b=7 {M_J}$, ${M_c}={M_d}=10 {M_J}$,
and ${M_e}=(7-10) {M_J}$, with semimajor axes of 
68, 38, 24, and 14.5 AU, respectively.
Figure from \cite{Marois10} with the permission of NPG. }%
\label{Imaging2}
\end{minipage}
\end{center}
\end{figure}

\section{Formation of Super-Earths}

Planet formation is one of the most outstanding problems in
astronomy. Despite centuries of theoretical efforts in explaining 
the formation of the planets of our solar system, this problem is still 
unresolved and the formation of planets is still an open question. 
Although it is 
widely accepted that planet formation begins by the coagulation of dust 
particles to larger objects in a circumstellar disk 
of gas and dust known as nebula, the details of this process are unknown 
and the formation of giant and terrestrial planets is not fully understood.

The issue is even more complicated in extrasolar planetary systems. 
The current models of planet formation, which have been developed
primarily for explaining the formation of the planets of our solar system, 
cannot explain the formation and dynamical diversity of many of 
extrasolar planets. The unexpected properties of these bodies have 
raised many questions about the validity of the current theories
of planet formation and their applicability to other planetary systems.
For instance, many of the currently known extrasolar giant planets
have orbits smaller than the orbit of Mercury around the Sun. This is an
anomaly that cannot be explained by the current theories of planet 
formation (as explained below, giant planets are expected to form
at large distances). The discovery of these {\it hot Jupiters} prompted
theoreticians to revisit models of giant planet formation, and attribute
the close-in orbits of these objects to their interactions with their
surrounding nebulae and their subsequent radial migrations
to closer orbits (Figure 6)\footnote{The resultant of the gravitational forces 
that a planet receives from the portions of the nebula interior and
exterior to its orbit causes the planet to radially migrate. 
The migration is classified as type I when the planet is small and
does not accrete nebular material (i.e. it does not create
a gap in the nebula while migrating). When the planet is large 
and accretes material during its migration, a gap will appear and the 
migration is classified as type II. See figure 6.}. We refer the reader to 
Chambers (2009; \cite{Chambers09}) and Armitage (2010; \cite{Armitage10})
for a comprehensive review of planet migration and its implications for the 
formation of planets.

\begin{figure}
\vskip 0.1in
\begin{center}
\center{
\resizebox*{7.5cm}{!}{\includegraphics{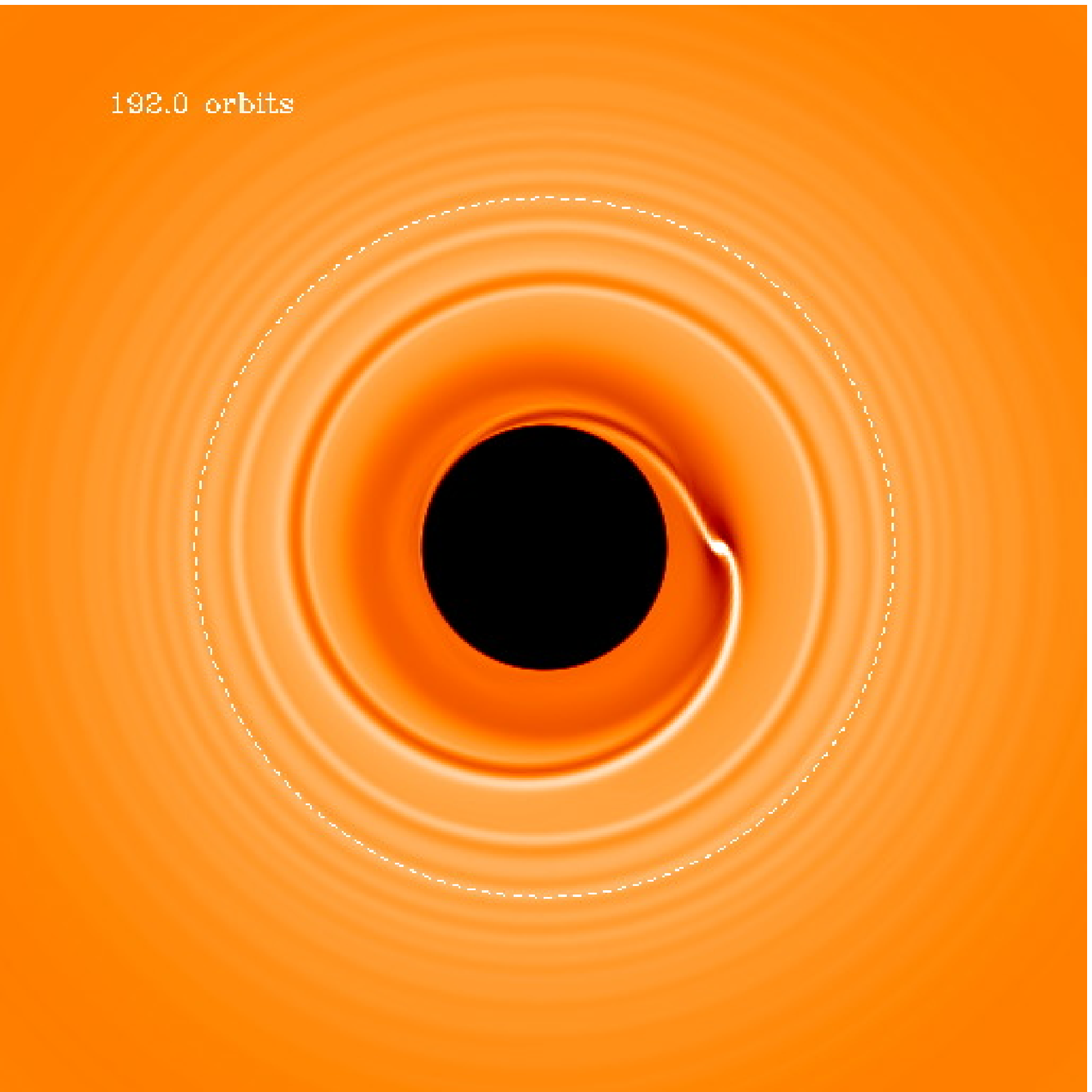}}
\resizebox*{7.5cm}{!}{\includegraphics{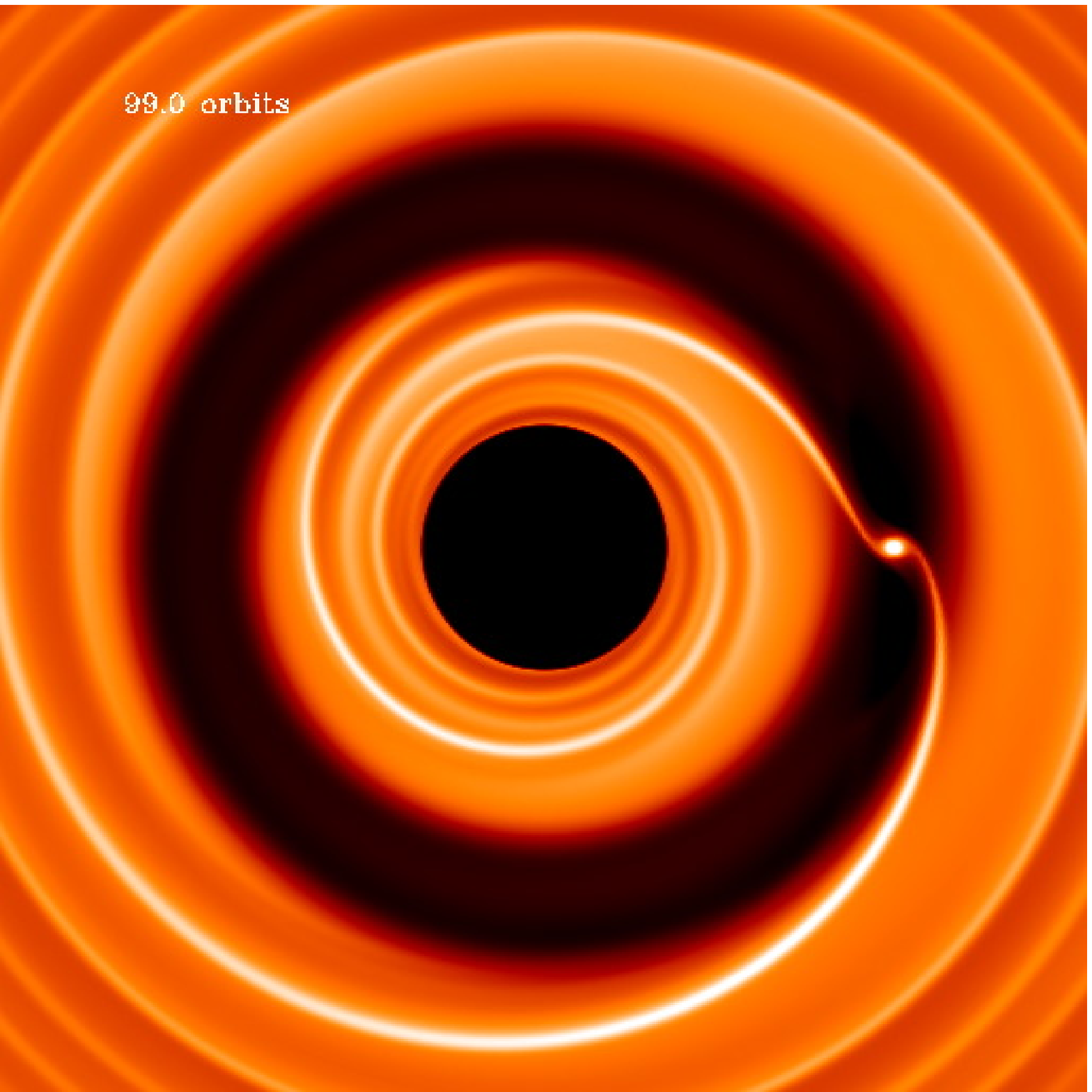}}}%
\vskip 10pt
\caption{Graphs of the type I (left) and type II (right) planetary migration.
In the left panel, the mass of the planet is small and no gap-opening occurs.
As the planet grows, it clears its surrounding and a gap
gradually appears. Figures courtesy of F. Masset.}%
\label{Planet-Migration}
\end{center}
\end{figure}

The existence of super-Earths is another of such unexpected findings.
While in our solar system, planets belong to two distinct categories of
terrestrial (i.e. Earth-sized or smaller) and giant (approximately 12 times more
massive than Earth and larger), super-Earths, with an intermediate mass-range,
introduce a new class of objects.
The planet formation theories not only have to now explain the formation of
these bodies; in some case, they also have to explain their unusual 
dynamical properties. 

This section focuses on the formation of super-Earths. 
It begins by explaining different models of the formation
of giant and terrestrial planets in our solar system, and discusses 
their applicability to the formation of super-Earth objects.

\begin{figure}
\vskip 0.1in
\begin{center}
\begin{minipage}{150mm}
\center{
\resizebox*{10cm}{!}{\includegraphics{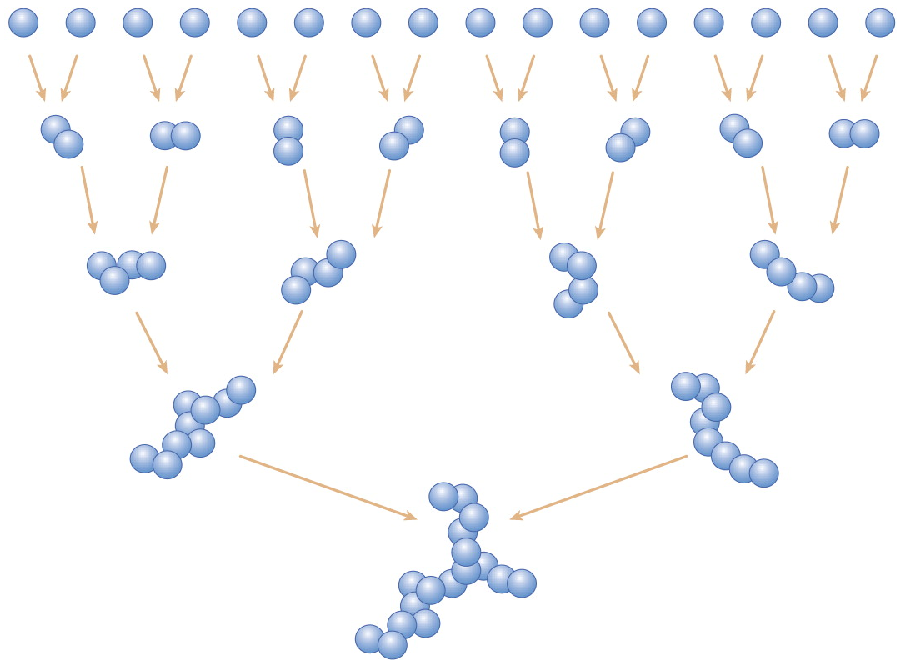}}%
\vskip 0.1in
\resizebox*{10cm}{!}{\includegraphics{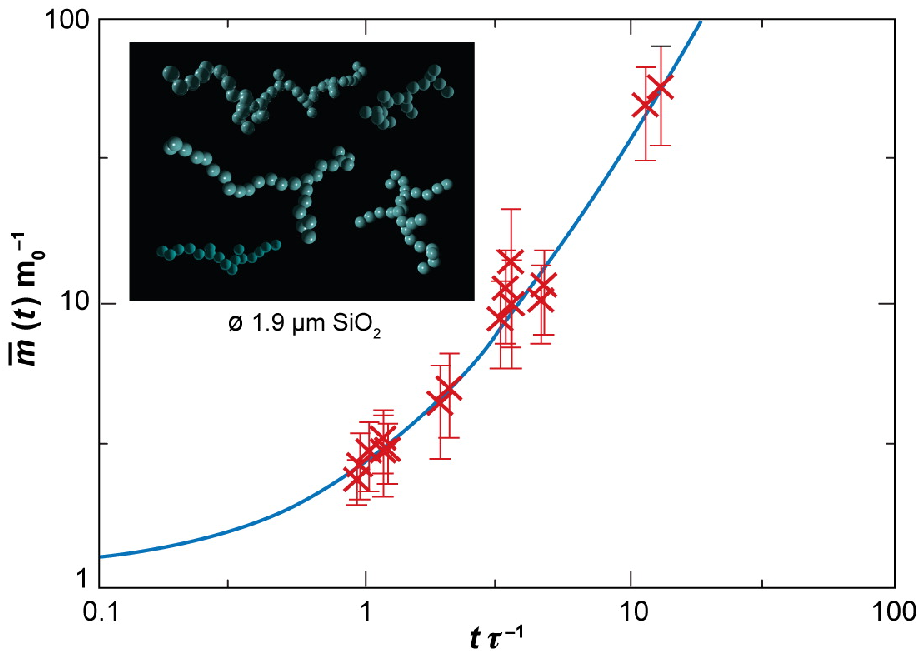}}}%
\caption{Top:  Hit-and-stick collisions lead to fractal growth and subsequently
to a narrow size distribution \cite{Blum06}. The collision among dust grains occurs
for several reasons including their Brownian motion in the gas. Bottom:
Graph of the fractal growth of aggregates during collisions in Brownian motion.
The data points are taken from experiments by Krause \& Blum 
(2004; \cite{Krause04}). The solid curve is the analytical model. The time on the horizontal axis
is normalized to the collisional timescale for grains. The inset on the upper left corner shows 
examples of fractal dust aggregates found in the space shuttle experiments by Blum et al
(2000; \cite{Blum00a}). Figure from \cite{Blum08} with the permission of ARAA.}%
\label{Dust-Growth}
\end{minipage}
\end{center}
\end{figure}

\begin{figure}
\begin{center}
\begin{minipage}{150mm}
\center{
\resizebox*{15cm}{!}{\includegraphics{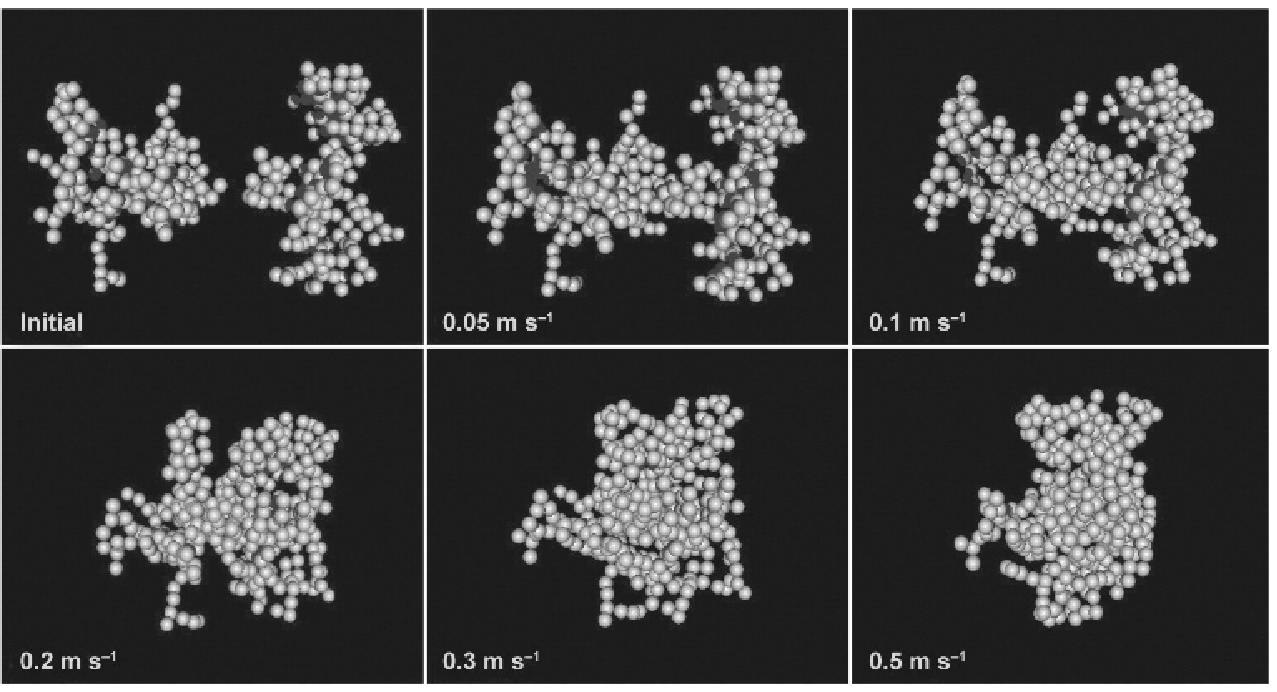}}}%
\caption{Molecular-dynamics simulations by Paszun \& Dominik (2006; \cite{Paszun06}) 
showing compaction of fractal dust aggregates.  The upper left panel shows 
two aggregates before collision. 
The merging and compaction of the aggregates are shown in subsequent panels.
Figure from \cite{Blum08} with the permission of ARAA.}%
\label{Fractal}
\end{minipage}
\end{center}
\end{figure}

\subsection{Models of Planet Formation}

Planets are formed in a circumstellar disk of gas and dust by
the coagulation of micron-sized dust grains to larger objects. 
In general, this process proceeds in four stages;

\begin{itemize}

\item growth of dust particles to centimeter- and decimeter-sized
bodies through gentle hitting and sticking,
\vskip 10pt
\item growth of centimeter- and decimeter-sized objects to km-sized
planetesimals,
\vskip 10pt
\item collisional growth of km-sized planetesimals to the cores of 
giant planets in the outer regions of the nebula, and to moon- and
Mars-sized bodies (known as protoplanets or planetary embryos) in the
inner regions,  and 
\vskip 10pt
\item the accretion of gas and formation of giant planets followed by
the collisional growth of planetary embryos to terrestrial bodies.

\end{itemize}

The first stage of this process is well understood. At this stage, 
dust grains are strongly
coupled to the gas and their dynamics is driven by non-gravitational
forces such as radiation pressure, and also by gas drag. Because dust particles follow
the motion of the gas, their relative velocities are small. As a result, they
slowly approach one another and gently collide. 
Laboratory experiments and computational simulations have shown that such gentle
collisions result in the fractal growth of dust grains to larger aggregates
(Figures 7 and 8)
\cite{Smoluchowski16,Dominik97,Blum98,Wurm98,Blum00,Krause04,Blum06,Wada07}.

The second stage (i.e. the growth of centimeter-sized objects to kilometer-sized
planetesimals) is still a big mystery. 
The collisions of centimeter- or decimeter-sized bodies with one another do 
not seem to facilitate the growth of these objects to larger sizes
\cite{Blum08,Wurm10}. As dust grains grow, their coupling to the gas weakens
(i.e., they move faster in the gas) \cite{Weidenschilling77}
and they show more of their independent dynamics. 
When two objects reach several centimeters or decimeters in size, their 
relative velocities become so large that their collisions may result 
in breakage and fragmentation (e.g., \cite{Wurm10}). Known as the 
{\it centimeter-sized barrier},
such disruptive collisions prevent the growth of small bodies to larger sizes.

The difficulties do not end here. In the event that some centimeter-sized
bodies manage to grow, the subsequent increase in their velocities
causes many of them, in particular those with sizes
of 1 to several meters, to either collide and shatter each other, or rapidly spiral 
towards the central star. Known as the {\it meter-sized barrier}, these effects 
deprive the nebula of enough materials to form planets.

The puzzling fact is that despite these difficulties, 
planets do exist and the above-mentioned issues were 
somehow overcome. 
Many theoretical models have been developed to solve this puzzle
\cite{Johansen07,Cuzzi08,Weidenschilling10}.
However, they all have limitations and none has been
able to present a complete and comprehensive scenario for the formation
of km-sized planetesimals. We refer the reader to articles by 
Blum \& Wurm (2008; \cite{Blum08}) and Chiang \& Youdin (2010, \cite{Chiang10}) 
for reviews of the current state of research in this area.

At the third stage of planet formation, the situation is different. 
Here, the interactions among planetesimals are primarily 
gravitational. Since the protoplanetary disk at this stage 
is populated by km-sized and larger objects, collisions among these bodies
are frequent. In general, frequent collisions in a crowded environment result 
in low eccentricities and low inclinations which facilitate the merging 
and accretion of the colliding bodies.
As a planetesimal grows, the influence zone of its gravitational field 
expands and it attracts more material from its surroundings. In other words, 
more material will be available for the planetesimal to accrete, and as a result
the rate of its growth is enhanced. Known as {\it runaway growth}, this process 
results in the growth of km-sized planetesimals to larger bodies in a short time
(Figure 9)
\cite{Safronov69,Greenberg78,Wetherill89,Wetherill93,Kokubo96,Ida93,Weidenschilling97,Kokubo00}.

\begin{figure}
\begin{center}
\begin{minipage}{150mm}
\center{
\resizebox*{7.5cm}{!}{\includegraphics{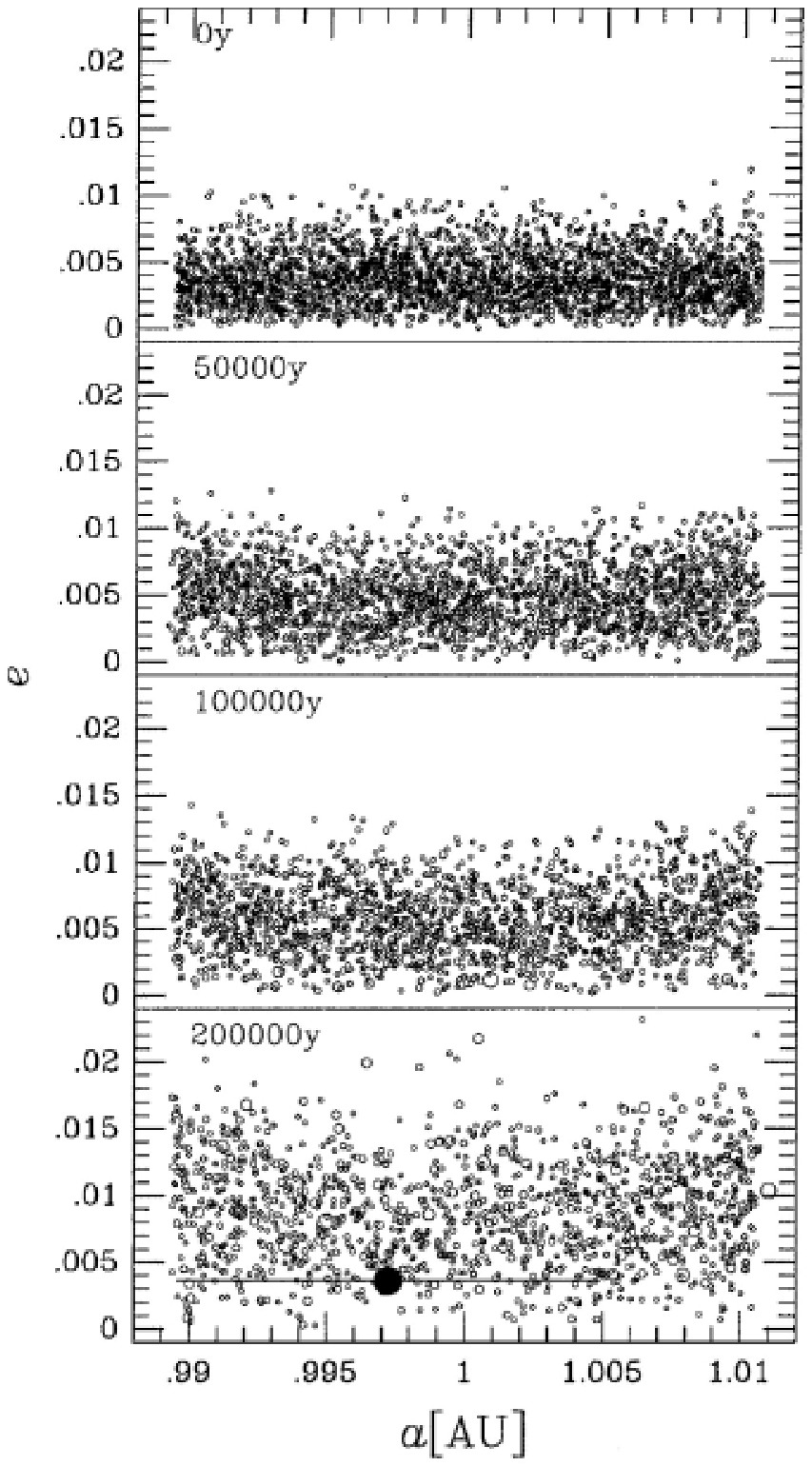}}%
\resizebox*{6.86cm}{!}{\includegraphics{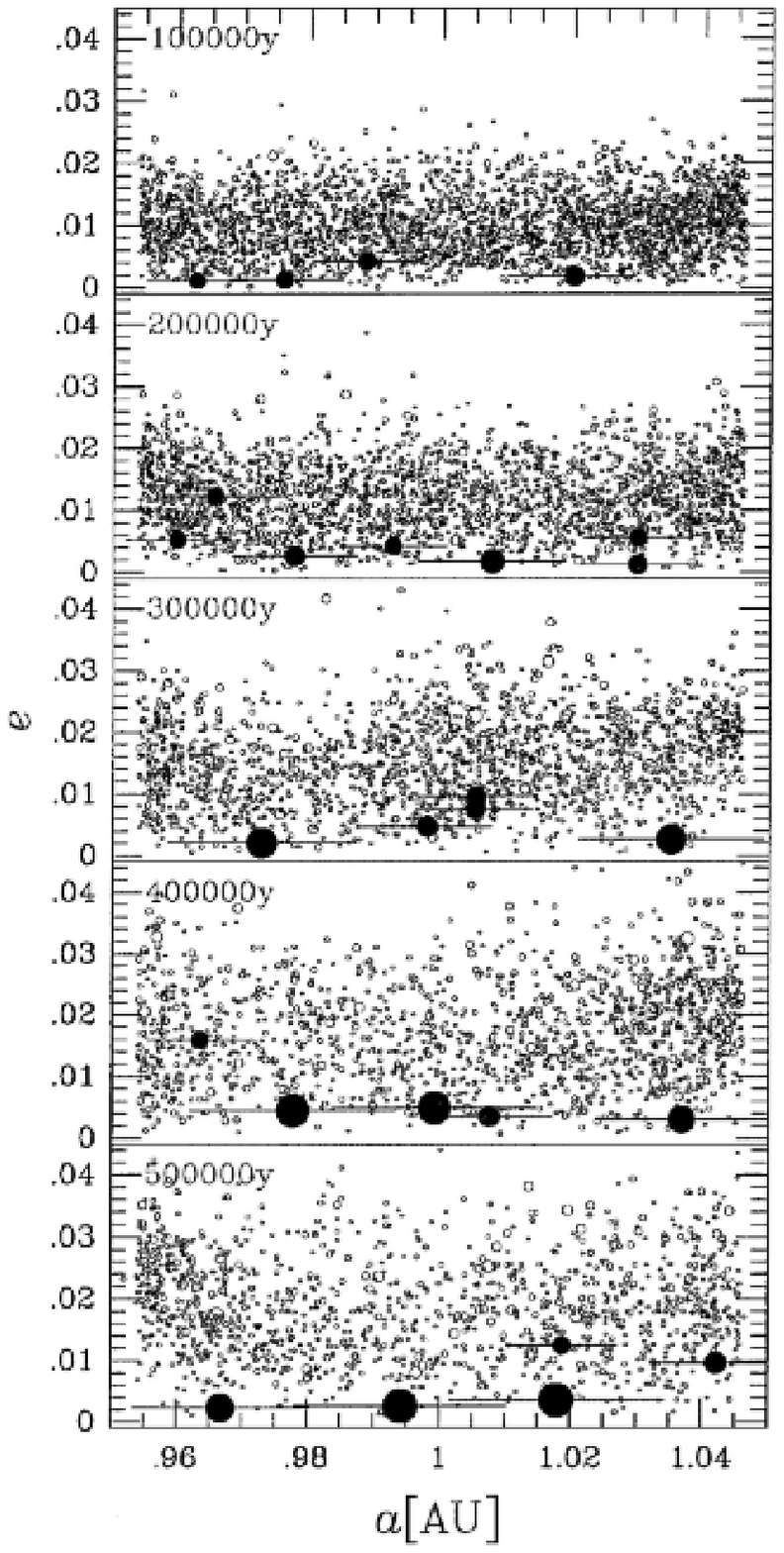}}}%
\caption{Snapshots of the evolution of a disk of planetesimals during their collisional growth
to larger objects. The left panel shows the formation of a protoplanet after $2 \times 10^5$
years. The right panel shows the interaction of protoplanets with planetesimals and the formation
of several moon-sized embryos. Figure from \cite{Kokubo00} with permission.}%
\label{Runaway-Growth}
\end{minipage}
\end{center}
\end{figure}

At large distances from the central star (e.g. $> 5$ AU from the Sun) 
where the rotational velocities are small, 
the collisional growth of planetesimals is more efficient. At such 
distances, planetesimals approach each other with small
relative velocities and their impacts are likely to result in accretion. Also, because 
far from the star, the temperature is low, the bulk material of such planetesimals 
is primarily ice which increases the efficiency of their sticking at the time 
of their collision. As a result, in a short time, planetesimals grow to large objects
with masses equal to a few masses of Earth. 
As this process occurs while the nebular gas is still around,
growing planetesimals gradually attract gas from their surroundings forming a large
body with a thick gaseous envelope and a mass equal to a few hundred Earth-masses.
At this state, a gas-giant planet is formed. 
This mechanism that is known as the 
{\it core-accretion} model is widely accepted as the model of the 
formation of gas-giant planets in our solar system (Figure 10)
\cite{Pollack96,Hubickyj05,Lissauer09,Movshovitz10}. 

At distances close to the central star, the accretion of planetesimals follows a
slightly different path.
Similar to the process of the formation of
the cores of gas-giant planets, the collisions of planetesimals 
at this stage may result in their growth to larger bodies. However, 
the efficiency of planetesimal accretion will not be as high
and as a result, instead of forming objects as big as the cores of giant planets,
accretion of planetesimals in this region results in
the formation of several hundred moon-sized bodies known as 
{\it planetary embryos}. Computational simulations \cite{Bromley06} and
analytical analysis \cite{Goldreich04} have shown that when the masses of these 
embryos reach the lunar-mass, planetesimals can no longer damp their orbits through 
dynamical friction, and the runaway growth ends. The gravitational perturbation of
the resulted planetary embryos affect the dynamics of smaller
planetesimals and cause them to
 collide with one another and/or be scattered to
large distances where they may leave the gravitational field of the system.
This growth and clearing process continues until
terrestrial planets are formed and the smaller remaining bodies (asteroids) 
are in stable orbits 
\cite{Kokubo95,Kokubo98,Morbidelli00,Chambers01,OBrien06a,Raymond06,Kokubo06,Kokubo07}.
Figure 11 shows the time evolution of a sample simulation of 
terrestrial planet formation \cite{Raymond06}.

\begin{figure}
\begin{center}
\center{
\resizebox*{7.5cm}{!}{\includegraphics{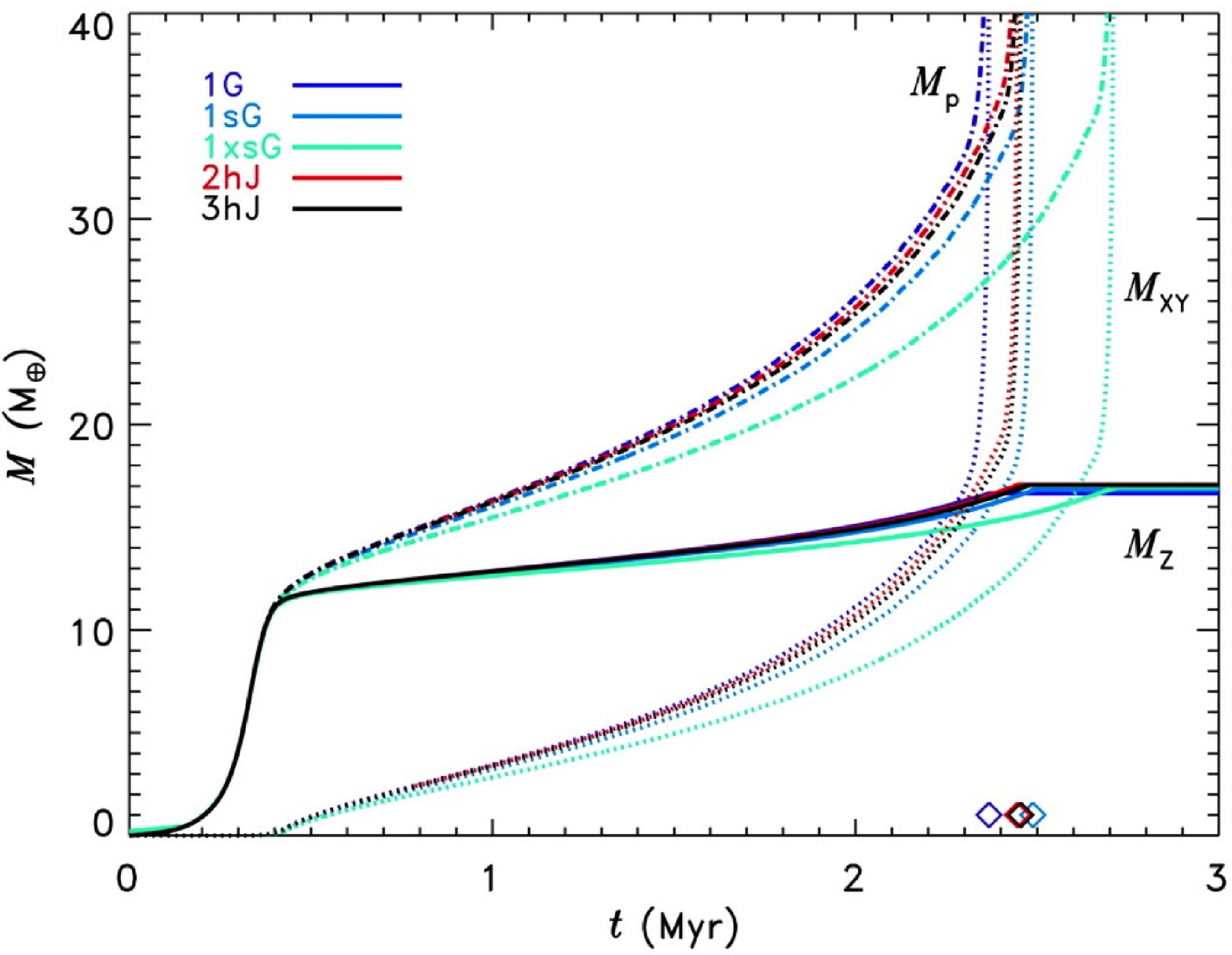}}
\resizebox*{7.5cm}{!}{\includegraphics{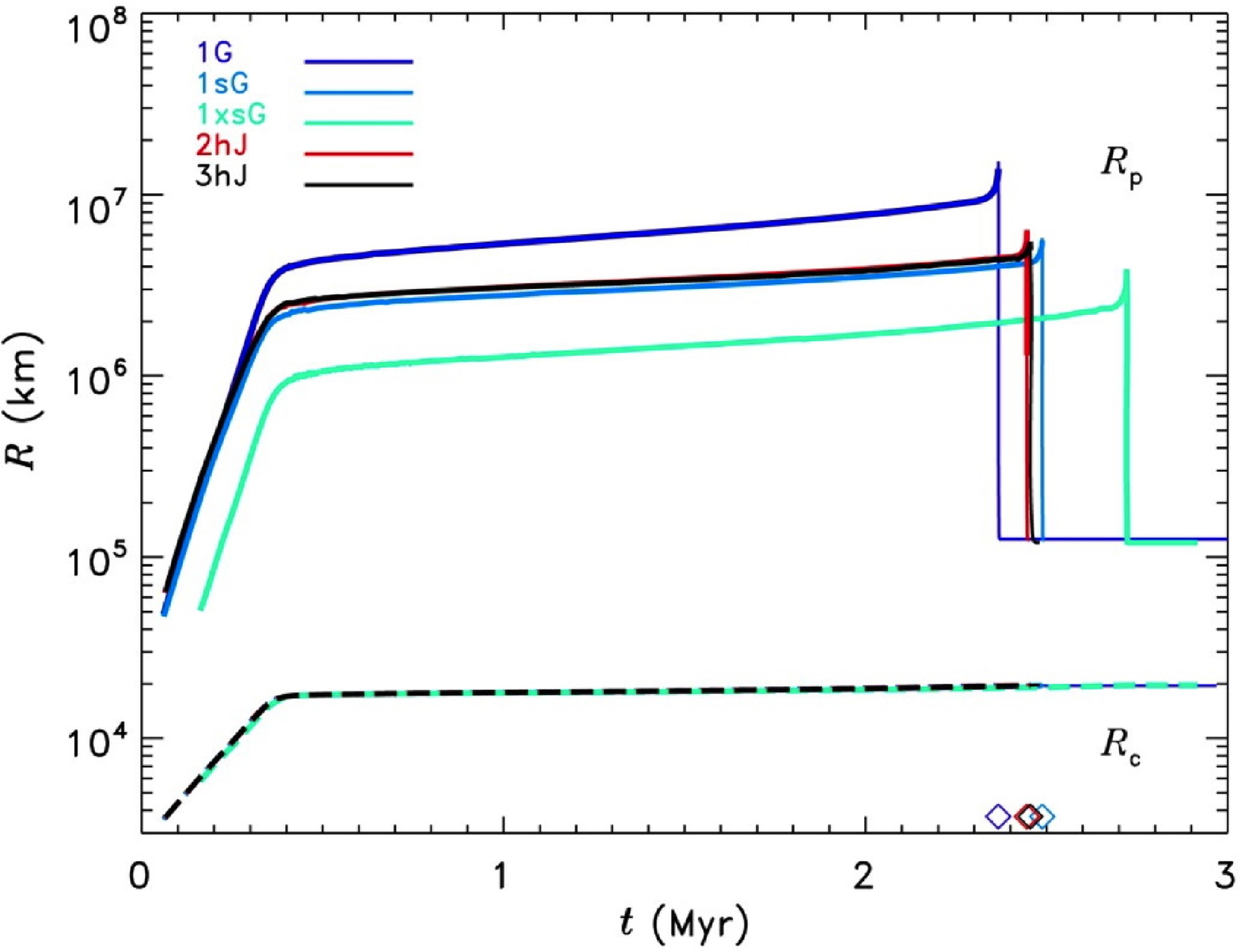}}
\vskip 10pt
\resizebox*{7.5cm}{!}{\includegraphics{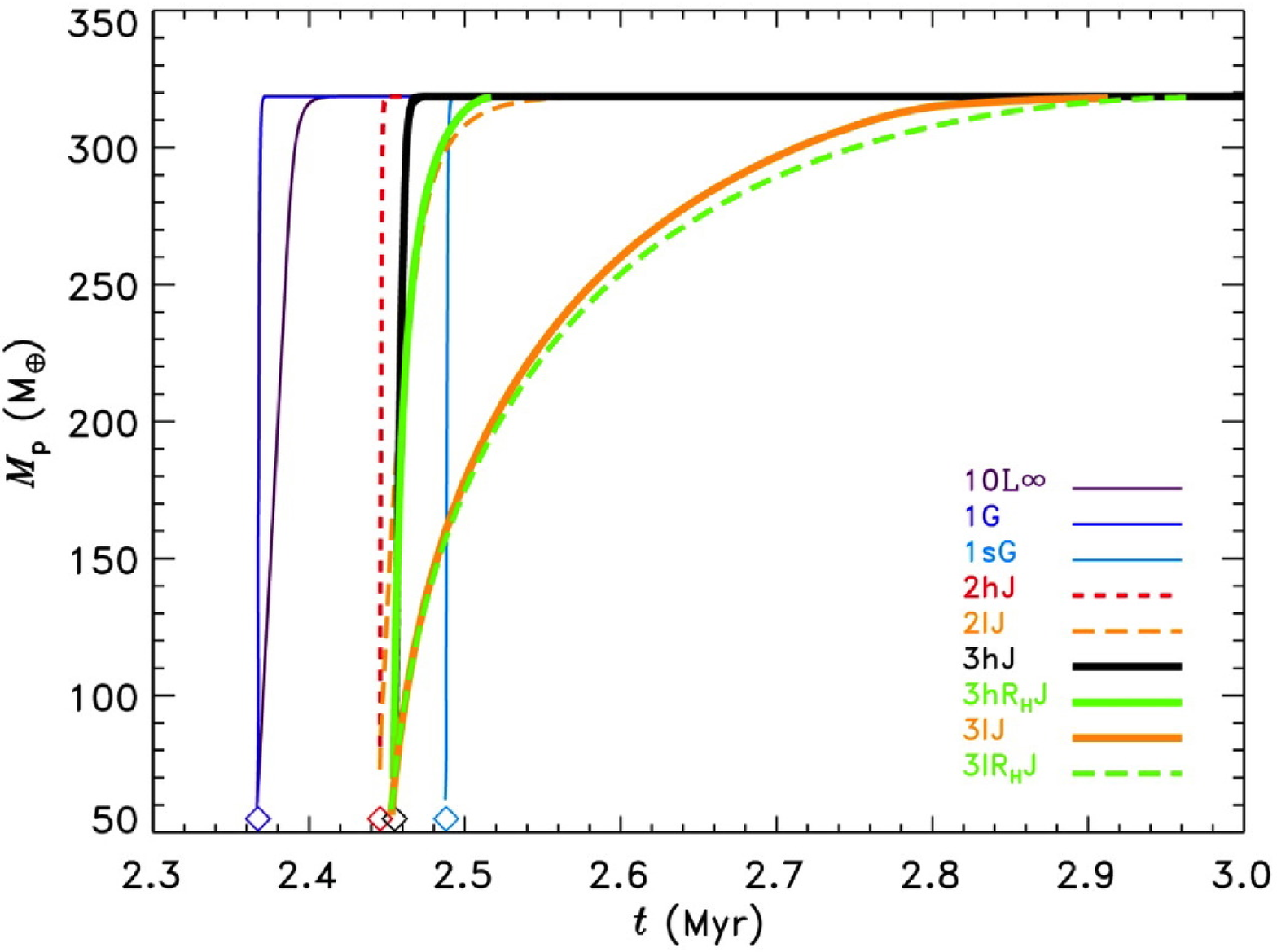}}
\resizebox*{7.5cm}{!}{\includegraphics{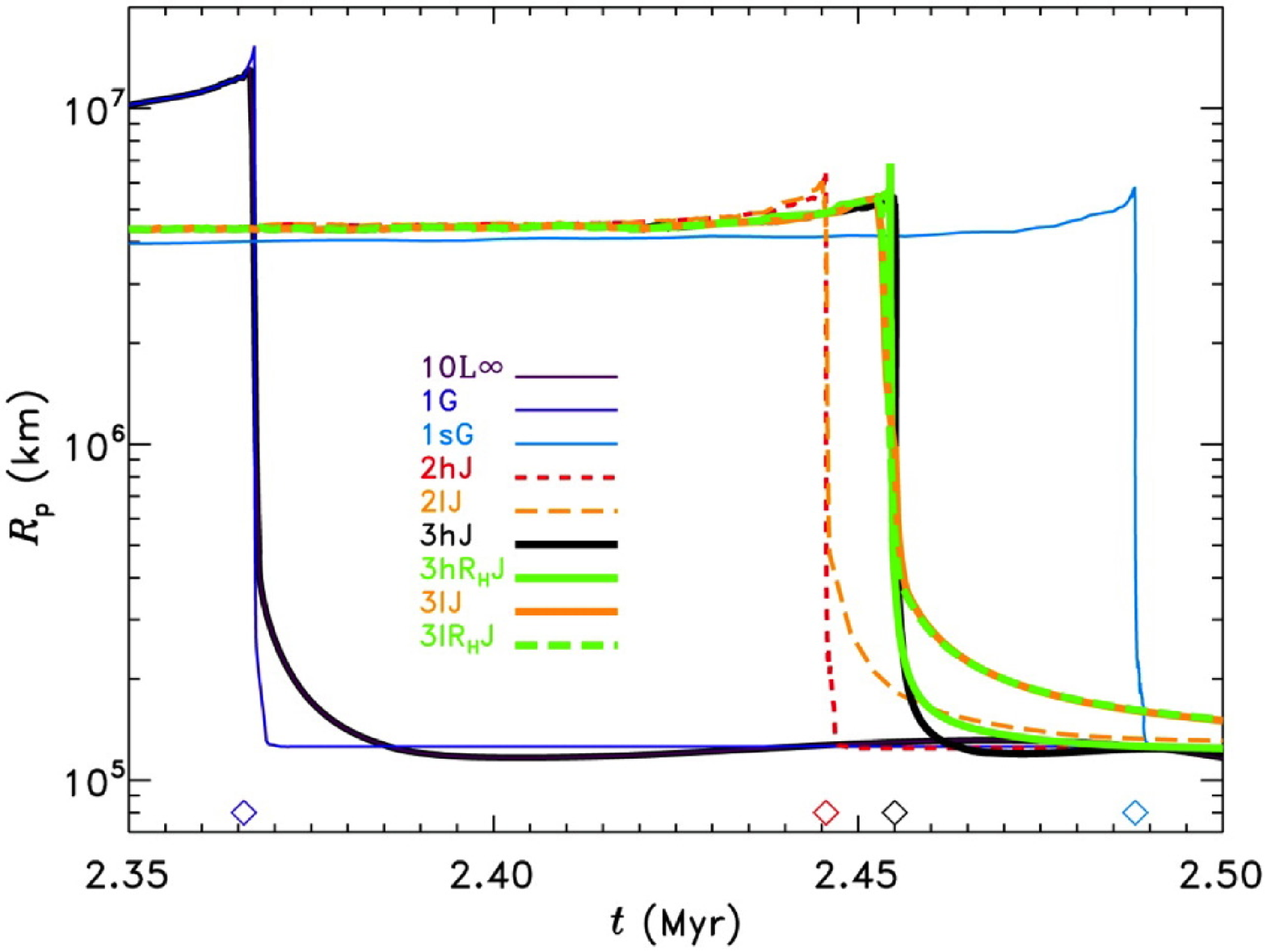}}}%
\caption{Graphs of the mass of the giant planet (left panels) and its radius (right panels)
for simulations by Lissauer et al (2009; \cite{Lissauer09}). 
Each color corresponds to a different
value of protoplanet surface density (see \cite{Lissauer09} for details).  The solid line 
represents the mass of the core, the dotted line shows that of the gaseous
envelope, and the dashes-dotted line corresponds to the total mass of the planet. The upper left 
graph shows the mass-growth only up to 40 Earth-masses in order to show the details 
of mass-accretion
at early stages. The top right panel represents the radius of the planet during this stage. 
The panel on the bottom left shows the total mass of the planet. As shown here, the giant planet
accretes more than 300 Earth-masses in approximately 3 Myr. The bottom right panel shows the
total radius of the planet. Figure from \cite{Lissauer09} with permission.}%
\label{Lissuer09}
\end{center}
\end{figure}

The above-mentioned processes, although seemingly straightforward, are
extremely complex. The immensity of the nebula, the enormous 
number of interacting objects,
and the complicated physics that is involved in their interactions make it impossible
for any simulation of planet formation to include all necessary components and to be
fully comprehensive. These simulations are also constantly 
challenged by observations that
reveal more characteristics of planet-forming environments. For instance, during
the formation of giant planets, the core-accretion model requires the nebular gas to
be available as the core of Jupiter grows and accrete gas from its surrounding.
The computational simulations presented in 
the original paper by Pollack et al (1996; \cite{Pollack96})
suggest that this time is approximately 10 Myr. 
In other words, in order for gas-giant planets to form by the core-accretion model,
the lifetime of the nebular gas has to be comparable with this time.
However, the observational estimates of the lifetimes of disks around young stars 
suggest a lifetime of 0.1-10 Myr, with 3 Myr being the age for
which half of stars show 
evidence of disks \cite{Strom93,Haisch01,Chen04,Maerker06}.
Any model of gas-giant planet formation has to be able to 
form these objects in less than approximately 3 Myr. 

Additionally, the simulations of the core-accretion model 
suggest that the core of Jupiter grows to $\sim 10$ Earth-masses.
However, computational modeling of the interior of Jupiter and Saturn point 
to values ranging from 0 to as large as 14 Earth-masses \cite{Guillot05,Militzer08}. 
It is unclear what the actual 
masses of the cores of our gas-giant planets are, and if smaller than 10 Earth-masses, how they 
accumulated their thick envelopes in a short time. We refer the reader to a review article
by Guillot (2005; \cite{Guillot05}) for more details.

\begin{figure}
\begin{center}
\center{
\resizebox*{12cm}{!}{\includegraphics{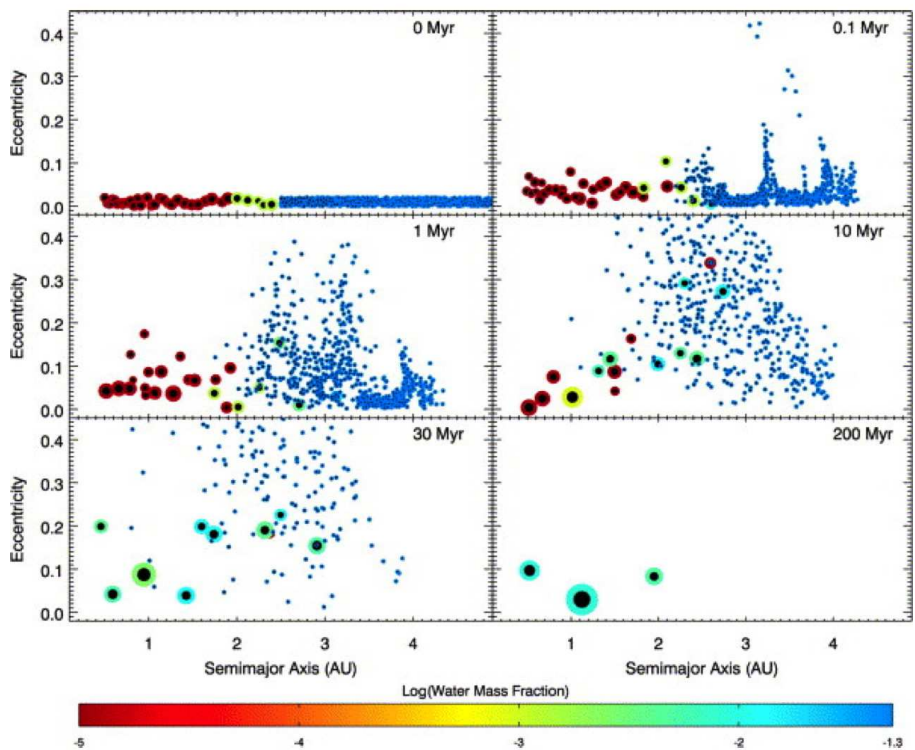}}}%
\caption{Radial mixing of planetesimals and planetary embryos, and the formation of
terrestrial planets. A Jupiter-sized planet, not shown in the figure, is at 5.5 AU. The size of
each object has been scaled with its mass assuming that it is a perfect sphere. The color of
each object corresponds to its water content (water/mass fraction). 
Red represents dry, light green represents 1\% water,
and blue corresponds to 5\% water content. 
The black circle in the middle of each object shows its
solid core. Figure from \cite{Raymond06} with permission.}%
\label{Water-delivery}
\end{center}
\end{figure}

Several efforts have been made to overcome these difficulties.
As shown by 
Hubickyj et al (2005; \cite{Hubickyj05}) and Lissauer et al (2009; \cite{Lissauer09}), 
increasing the surface density of the nebula to higher than that suggested by
Pollack et al (1996; \cite{Pollack96}) significantly reduces the time of the giant planet 
formation (Figure 10). 
Furthermore, an improved treatment of the grain physics as given by 
Podolak (2003; \cite{Podolak03}), Movshovitz \& Podolak (2008; \cite{Movshovitz08}), 
and Movshovitz et al (2010; \cite{Movshovitz10})
indicates that the value of the grain opacity in the envelope of the growing Jupiter
in the original core-accretion model \cite{Pollack96} is too high, and a lower value has to be 
adopted. This lower opacity has led to a revised version of the 
core-accretion model in which the time of giant planet formation 
is considerably smaller \cite{Hubickyj05,Movshovitz10}. 
Most recently, Bromley \& Kenyon \cite{Bromley11} have developed a new 
hybrid N-body-coagulation code which enables this authors to form Saturn- and
Jupiter-sized planets in approximately 1 Myr.

An alternative mechanism, known as the {\it disk instability} model, addresses this
issue by proposing rapid formation of giant planets in a gravitationally unstable nebula
\cite{Boss00a,Boss00b,Mayer02,Boss03,Mayer04,Durisen07,Mayer07,Boley09,Boley10,Cai10}. 
In this model, local gravitational 
instabilities in the solar nebula may result in the fragmentation of the disk to massive 
clumps which subsequently contract and form gas-giant planets in a short time (Figure 12).
Calculations by Boss (2000; \cite{Boss00a,Boss00b}) and 
Mayer et al (2002-4; \cite{Mayer02,Mayer04})
show that an unstable disk can break up into giant gaseous 
protoplanets in approximately 1000 years. Although this mechanism presents a fast track 
to the formation of a gas-giant planet, it suffers from the lack of an efficient cooling 
process necessary to take energy away from a planet-forming
clump in a sufficiently short time before it disperses.

\begin{figure}
\begin{center}
\center{
\resizebox*{13cm}{!}{\includegraphics{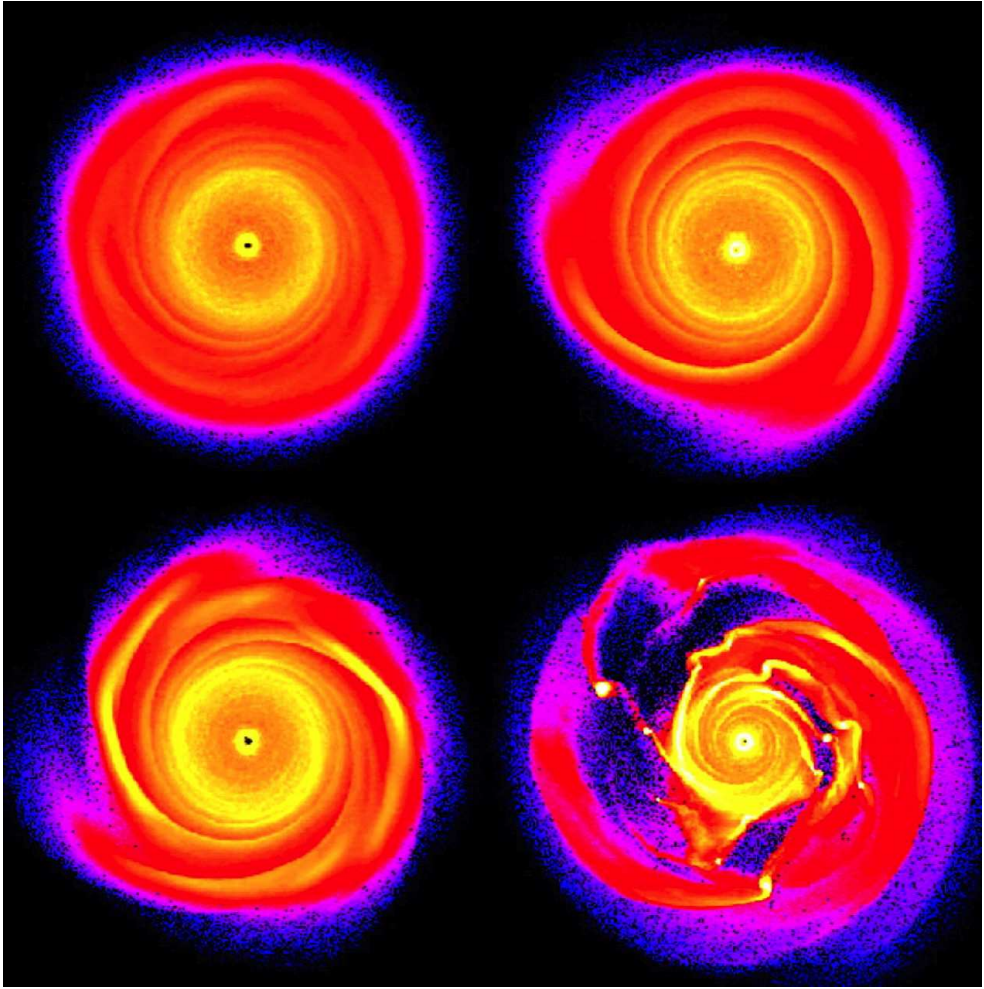}}}%
\caption{Snapshots of the evolution of a protoplanetary disk in the disk instability model.
As shown here, while the disk evolves, spiral arms appear where the density of the gas is
locally enhanced (bright colors correspond to high densities), and clumps are formed.
Figure from \cite{Mayer02} with the permission of AAAS.}%
\label{Disk-Instability}
\end{center}
\end{figure}

\subsection {Application to the Formation of Super-Earths}

As explained above, the current models of giant and terrestrial planet 
formation have been developed to explain the formation of the 
planets in our solar system. Since there are no super-Earths around 
the Sun, it may not be obvious whether these models can 
also explain the formation of these objects. However, a deeper 
look at the ranges of the masses and sizes of these bodies suggests
that super-Earths might have formed in the same way as gas-giant planets.
The key is in the intermediate range of the masses
of these objects. With masses ranging from a few Earth-masses to slightly smaller
than Uranus,
super-Earths are basically as massive as the cores of gas-giant planets. 
In fact some researchers consider super-Earths as giant planets'
``failed cores''. We recall that according to the core-accretion model and
the simulations of the interior of Jupiter, this planet
may have a core with a mass between zero and 14 Earth-masses \cite{Guillot05,Militzer08}.

The extent to which the current models of giant planet formation 
can be used to explain the formation of 
super-Earths varies from one system to another.
The diversity of the currently known super-Earth planetary 
systems, both in spectral types of their host stars and the 
orbital dynamics of their planets (see Table 1), suggests that while
in some systems (e.g., around M stars)
super-Earths might have formed in-place 
\cite{Laughlin04,Kennedy06,Kennedy07,Kennedy08a,Kennedy08b,Ida08}, 
in other systems (e.g., around G stars) the formation of these objects 
might have occurred while their orbital elements were evolving 
\cite{Terquem07,Kennedy08a,Kennedy08b}. In such systems, the larger than terrestrial 
masses of super-Earths, 
combined with the fact that many of these objects are in short-period orbits, 
points to a formation scenario in which
super-Earths were formed at large distances (where more material was available
for their growth) and either were scattered to their current orbits
as a result of interactions with other cores and/or planets \cite{Terquem07}, or migrated 
to their current locations as they interacted with the protoplanetary disk 
\cite{Kennedy08b}. 
This mechanism naturally favors the core-accretion model of gas-giant planet
formation, although attempts have also been made to explain the formation
of super-Earths via the disk instability scenario \cite{Boss06}.

\begin{figure}
\begin{center}
\center{
\resizebox*{12cm}{!}{\includegraphics{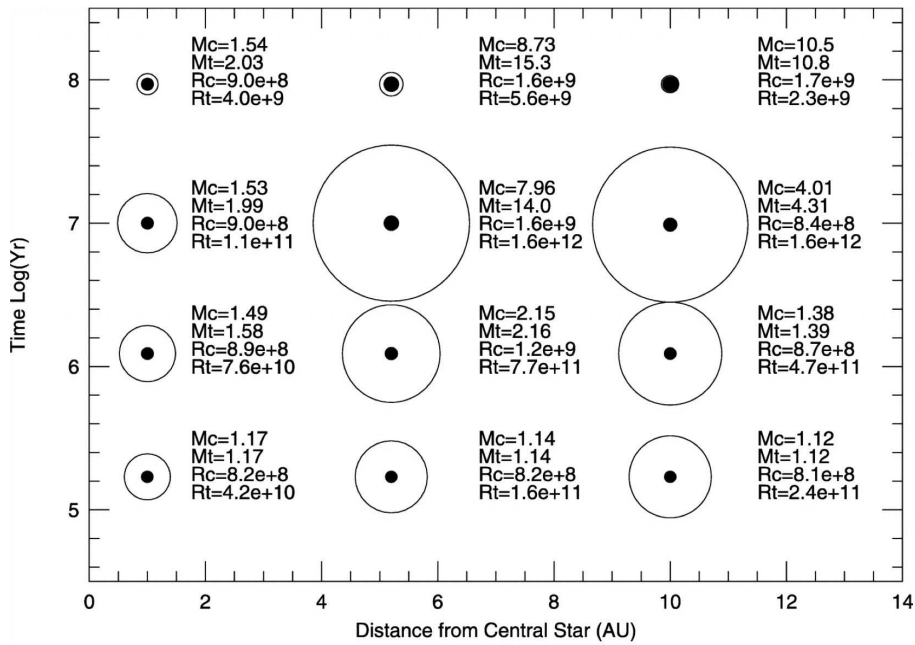}}}%
\caption{Graph of the core and envelope of a planet at different distances in a disk
around a 0.4 solar-mass star. Black circles represent cores and the large circles
are their envelopes. The mass of the core $({M_c})$ is given in Earth-masses and 
its radius $({R_c})$ is in centimeters. The total mass and radius of the planet 
are shown by ${M_t}$ and ${R_t}$, respectively. As shown here, in about 100 Myr,
the planet's envelope contracts and the planet reaches its final size.
Figure from \cite{Laughlin04} with the permission of AAS.}%
\label{Laughlin04-I}
\end{center}
\end{figure}

\vskip 10pt
\subsubsection{Formation of Super-Earths Around Low-Mass Stars: Core-Accretion}

To determine whether super-Earths can form in-place around low-mass stars,
Laughlin et al (2004; \cite{Laughlin04}) simulated giant planet
formation through the core-accretion model in disks around stars with 
masses smaller than 0.5 solar-masses. These authors showed that around M stars, this 
mechanism produces planets ranging from terrestrial-class 
to Neptune-sized (Figure 13). The results of their simulations 
also indicated that the lower-than-solar masses of M stars
(typically smaller than 0.4 solar-masses), which implies low masses and 
surface densities for their circumstellar disks as well,
results in less frequent collisions among planetesimals and planetary 
embryos, and prolongs the growth of these objects to larger sizes.
Consequently, the time of the core growth around M stars will be several 
times longer than the time of the formation of Jupiter around the Sun.
During this time, the gaseous component of the circumstellar disk 
is dispersed, leaving the slowly growing core with much less gas to accrete. 

The short lifetime of the gas in circumstellar disks around
M stars can be attributed to two factors: 1) the high internal 
radiation of young M stars (these stars are almost as bright as 
solar-type stars), and 2) external perturbations from other close-by stars.
Since most stars are formed in clusters \cite{Lada03},
their circumstellar disks are strongly affected by the gravitational
perturbations and the radiations of other stars \cite{Adams04}.
For M stars, this causes the 
circumstellar disk to receive high amount of radiation from both the central star
and external sources. The high amounts of radiation combined with the low masses of M stars,
which points to their small gravitational fields, increases the 
effectiveness of the photo-evaporation of the gaseous component of the 
circumstellar disk by up to two orders of magnitude. As a result, the majority of the
gas leaves the disk at the early stage of giant planet formation.

The slow growth of planetary embryos around M stars and the rapid dispersal of the nebular gas 
suggest that no giant planet should exist in these systems and the 
planets around M stars have to be mainly super-Earths or smaller. 
While the observational evidence is in agreement with the latter (e.g., the M star
GL 581 is host to 4-6 planets with masses similar to that of Neptune and smaller), it does not
support the first suggestion. Several giant planets have in fact been discovered 
around M stars among
which one can name GJ 876 with two planets with masses of 0.6 and 1.9 Jupiter-masses 
on 30-day and 60-day orbits \cite{Rivera10}, and HIP 57050 with a Saturn-mass planet in
its habitable zone \cite{Haghighipour10}.

\begin{figure}
\begin{center}
\center{
\resizebox*{11cm}{!}{\includegraphics{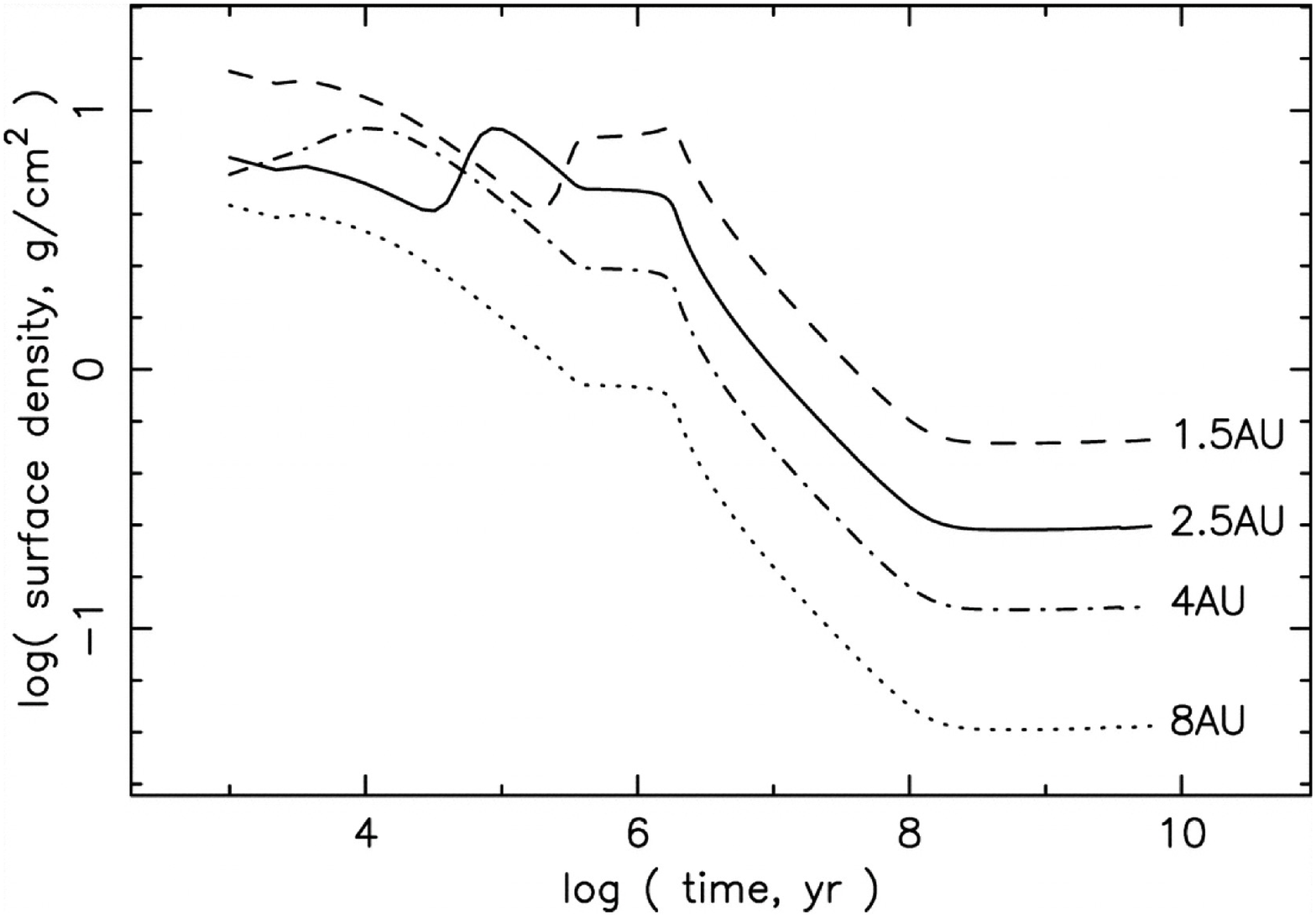}}
\vskip 10pt
\resizebox*{11cm}{!}{\includegraphics{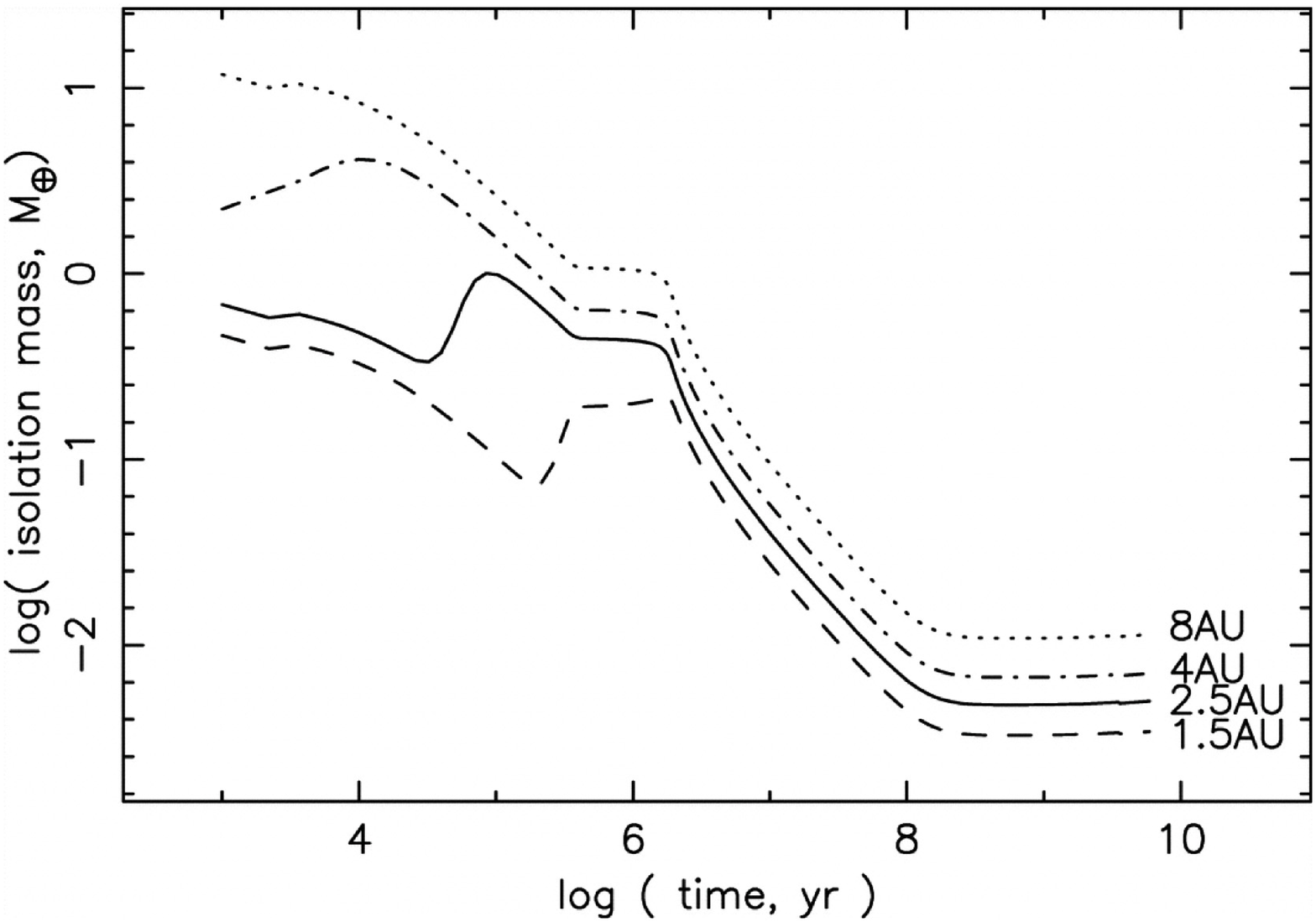}}}%
\caption{Top: The graph of the time evolution of the surface density of a disk around
a 0.25 solar-mass star. The initial mass of the disk is 0.065 times the mass of the
central star. Each curve on this graph corresponds to the disk evolution at that fixed
radius. Bottom: The graph of the mass of a planetary embryo at the indicated distance 
during the evolution of the disk. As shown by the two panels, the inward migration of 
snow line increases ice condensation which in turn results in an
increase in the disk surface density, and formation of larger objects. The latter is
more pronounced in the region between 2 and 8 AU. Figure from \cite{Kennedy06} 
with the permission of AAS.}%
\label{Kennedy06}
\end{center}
\end{figure}

\vskip 20pt
\noindent
{\em Effect of Stellar Evolution}
\vskip 5pt
The above-mentioned simulations do not take
the effect of stellar evolution
into account. As opposed to young solar-type stars whose luminosities stay almost
constant during the formation of a planet (e.g. 10 to 100 Myr),
the luminosity of a pre-main sequence low-mass star 
(e.g., 0.5 solar-masses) fades by a factor of 10 to 100 during this time
\cite{Hayashi81}. As a result, the internal temperature of the circumstellar disk
will decrease which causes the region known as 
{\it snow line} (or ice condensation limit, the region
beyond which water is in the permanent state of ice)
to move to close distances. The inward migration of the snow line 
results in an increase in the population of icy materials
(km-sized and larger planetesimals), which in turn
increases the efficiency of the collisional growth of these objects
to protoplanetary bodies (we recall that as mentioned in section 2.1, 
sticking is more efficient among icy bodies). 
As shown by Kennedy et al (2006; \cite{Kennedy06}),
around a 0.25 solar-mass star, the moving snow line causes
rapid formation of planetary embryos within a few million years.
Subsequent collisions and interactions among these objects result
in the formation of super-Earths in approximately 50--500 million years (Figure 14).

\vskip 20pt
\noindent
{\em Effect of Planet Migration}
\vskip 5pt
A common feature among the formation scenarios mentioned above is
the implicit assumption that planets are formed in-place. Although the post-formation
migration has been presented as a mechanism for explaining the close-in orbits
of super-Earths, these scenarios do not include the effect of the possible migration 
of still-forming planets (for instance, at the stage when cores of 
giant planets are forming) on the collisional growth of protoplanetary bodies. 
They also do not consider the possibility of the
migration of planetary embryos during the accretion of these objects.
However, studies of the interactions of disks and planets
have made it certain that planet migration occurs and has
profound effects on the formation of planetary systems and 
the final assembly of their planets and smaller constituents. 

In our solar system, planetary and satellite migration has long been recognized as a major
contributor to the formation and orbital architecture of planets, their moons, and other 
minor bodies.
For instance, as shown by Greenberg (1972-3; \cite{Greenberg72,Greenberg73}), 
mean-motion resonances (i.e., commensurable orbital periods\footnote{It is necessary to 
emphasize 
that orbital commensurability is necessary for two planets to be in a mean-motion resonance, 
however it is not sufficient. Other constraints have to exist between the angular elements of 
their orbits as well. For more details, the reader is referred to books on celestial mechanics.}) 
among the natural satellites of giant planets (e.g., Titan and Hyperion, satellites of Saturn) 
might have been the results of the radial migration 
of these objects due to their tidal interactions with their parent 
planets \cite{Goldreich65}.
Similarly, the orbital architecture of Galilean satellites and their capture in
a three-body resonance has been attributed to the migration of these objects first
during their formation while interacting with Jupiter's circumplanetary disk of satellitesimals
\cite{Canup02}, and subsequently by tidal forces after their formation \cite{Peale02}. 
The lack of irregular satellites between Callisto, the outermost Galilean
satellite, and Themisto, the innermost irregular satellite of Jupiter can also be explained
by a dynamical clearing process that occurred during the formation and migration of Galilean 
satellites \cite{Haghighipour08a}.

\begin{figure}
\begin{center}
\center{
\resizebox*{12cm}{!}{\includegraphics{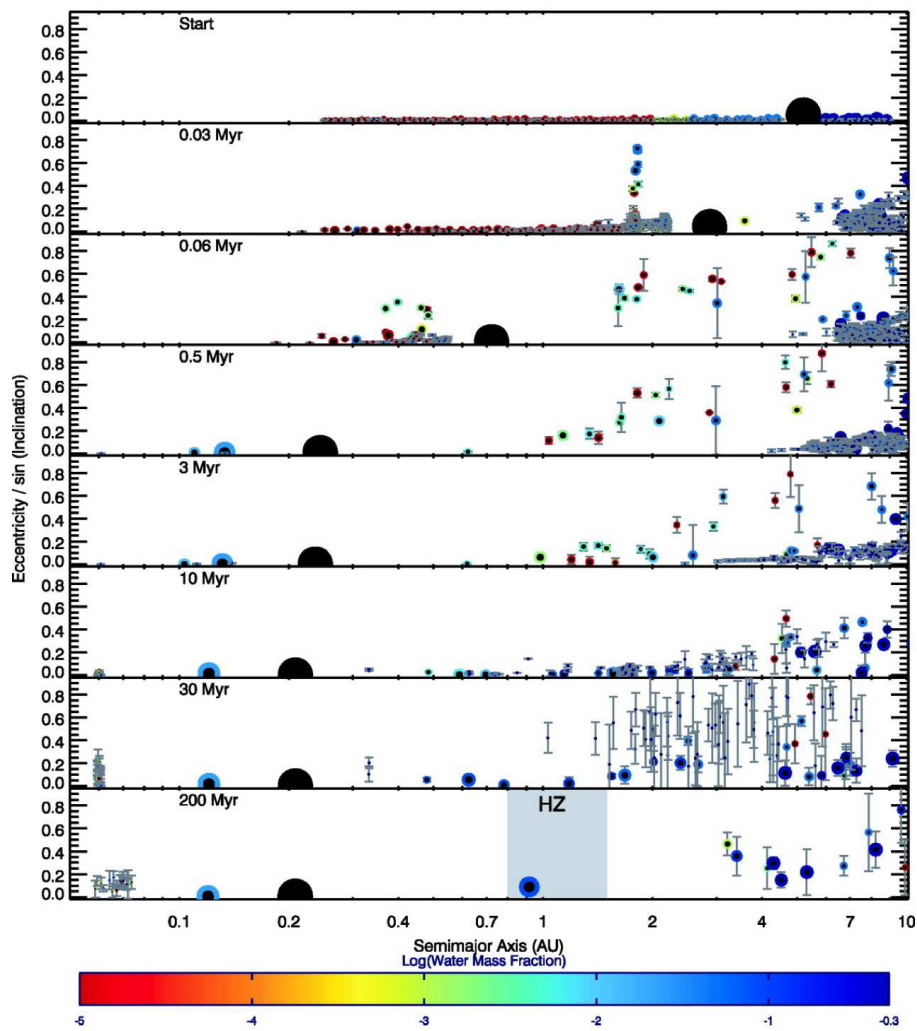}}}%
\caption{Graph of the formation of terrestrial planets during the migration
of a Jupiter-sized body. Colors indicate water content as in figure 7. Simulations
include gas drag as well. As shown here, while the giant planet migrates inwards, 
many protoplanetary bodies are either scattered out of the system, or their
eccentricities are lowered due to the gas drag, and they stay in orbits at larger 
distances. The collisions among the latter embryos may result in the formation
of terrestrial planets beyond the orbit of the giant body.
Figure from \cite{Mandell07} with the permission of AAS.}%
\label{Mandell07}
\end{center}
\end{figure}

The idea of the migration of planetary bodies was first proposed by 
Fernandez \& Ip (1984; \cite{Fernandez84}). These authors 
suggested that after the dispersal of the nebular gas, giant planets 
may drift from their original orbits due to the exchange of angular momentum with the 
planetesimal debris disk, and scatter these objects to other regions of the solar system.
This idea was later utilized by Malhotra (1993-5; \cite{Malhotra93,Malhotra95}) 
to explain the peculiar 
(highly eccentric, inclined, and long-term chaotic) orbit of Pluto, and by 
Malhotra (1996; \cite{Malhotra96}), and Hahn \& Malhotra (2005; \cite{Hahn05}) 
to explain the dynamical structure of Kuiper belt objects. 

Planetary migration has been used extensively to explain the existence of close-in 
Jupiter-like planets. 
In fact, it was the detection of the first hot Jupiter around the star 51 Pegasi \cite{51Peg}
that prompted scientists to revisit theories of planet migration in our solar system,
and apply them to extrasolar planets.
At present, planet migration is well-developed and widely accepted as part of 
a comprehensive formation mechanism. As mentioned in the Introduction, depending 
on the physical and
dynamical characteristics of planets and their circumstellar disks, migration occurs
in different forms (e.g., Type I and Type II, see Figure 6). 
Numerous articles have 
been published on this subject which unfortunately makes it impossible to cite 
them all here. We refer the reader to papers by Nelson et al (2001;\cite{Nelson01}), 
Masset \& Snellgrove (2001; \cite{Masset01}), Papaloizou \& Terquem (2006; \cite{Terquem06}) 
and articles by Chambers (2009; \cite{Chambers09}) and Armitage (2010; \cite{Armitage10})
for a review on this topic and
the effects of planet migration on the formation and dynamical evolution
of planetary systems.

The contribution of planet migration to the formation of close-in super-Earths 
may appear in different ways. A fully formed migrating giant planet affects
the dynamics of interior protoplanetary bodies by either increasing
their orbital eccentricities and scattering them to 
larger distances, or causing them to migrate to closer orbits. The migrated 
protoplanets may be {\it shepherded} by the giant planet into small close-in 
regions where they are captured in mean-motion resonances. As shown by 
Zhou et al (2005; \cite{Zhou05}),
 Fogg \& Nelson (2005-9; \cite{Fogg05,Fogg06,Fogg07a,Fogg07b,Fogg09}), 
and Raymond et al (2008; \cite{Raymond08}), around Sun-like stars, the shepherded 
protoplanets may also collide and grow to terrestrial-class and super-Earth objects. 
Also, as shown by Mandell \& Sigurdsson (2003; \cite{Mandell03}), 
Raymond et al (2006; \cite{Raymond06a}),  and Mandell et al (2007; \cite{Mandell07}),
in more massive protoplanetary disks around such stars, the out-scattered 
protoplanets may collide and grow to planetary-sized bodies (Figure 15).

Recent simulations by Haghighipour \& Rastegar (2010; \cite{Haghighipour10c}) have
shown that such accretion of protoplanets during giant planet migration may not be
efficient around low-mass stars. Simulating the dynamics of protoplanetary bodies
at distances smaller than 0.2 AU around a 0.3 solar-mass star, these authors have
shown that during the inward
migration of one or several giant planets (the latter involves migrating planets
in mean-motion resonances),
the majority of the protoplanets leave the system and do not contribute
to the formation of close-in Earth-sized and/or super-Earth bodies. Their results 
suggest that the currently known small planets around M stars 
might have formed at larger distances and were either
scattered to their current close-in orbits, or migrated into their orbits while
captured in mean-motion resonance with a migrating planet.

In a protoplanetary disk, the interactions among
cores of giant planets and planetary embryos may also result in 
the inward migration of the latter objects. While migrating, orbital crossing
and collisional merging of these bodies may result in their growth to
a few super-Earths, especially in mean-motion resonances. Simulating the
interactions of 25 protoplanetary objects with masses ranging 
from 0.1 to 1 Earth-masses, Terquem \& Papaloizou (2007; \cite{Terquem07}) 
have shown that a few close-in super-Earths may form in this way
with masses up to 12 Earth-masses. The results of the simulations 
by these authors suggest
that in systems where merging of migrating cores results in the formation
of super-Earth and Neptune-like planets, such planets will always be
accompanied by giant bodies and most likely in mean-motion resonances.
Similar results have also been reported by 
Haghighipour \& Rastegar (2010; \cite{Haghighipour10c}).

Interestingly, unlike the scenarios explained above, there are several planetary 
systems that host small Naptune-sized objects and super-Earths but do not harbor giant 
planets (e.g., HD 69830, GL 581). 
The planets in these systems do not have a Jupiter-like companion that
might have facilitated their formation. Such systems
seem to imply that a different mechanism may be responsible for the
formation of their super-Earth objects.
Kennedy \& Kenyon (2008; \cite{Kennedy08b}) and  Kenyon \& Bromley (2009; \cite{Kenyon09}) 
suggested that the migration of protoplanetary embryos may be the key in
facilitating the close-in accretion of 
these bodies. Similar to giant planets, planetary embryos can also
undergo migration. Simulating the growth 
of planetary embryos in a circumstellar disk with a density enhancement
at the region of its snow line, these authors have shown that
during the collision and growth of planetary embryos,
many of these objects may migrate towards the
central star. Around a solar-type star, the time of such migrations for an Earth-sized
planet at 1 AU is approximately ${10^5}$--${10^6}$ years--much smaller than the time of the 
chaotic growth of a typical moon- to Mars-sized embryos 
($10^8$ years) \cite{Goldreich04}. This
implies that most of the migration occurs prior to the onset of 
the final growth. Depending on their relative velocities,
the interaction of the migrated embryos may result
in the growth, scattering, and shepherding, as in the case of a
migrating giant planet. Simulations by 
Kennedy \& Kenyon (2008; \cite{Kennedy08b}) and  Kenyon \& Bromley (2009; \cite{Kenyon09}) 
have shown that super-Earth objects with masses up to
8 Earth-masses may form in this way around stars ranging from 0.25 to 2 solar-masses 
(Figure 16).

\begin{figure}
\begin{center}
\center{
\resizebox*{10cm}{!}{\includegraphics{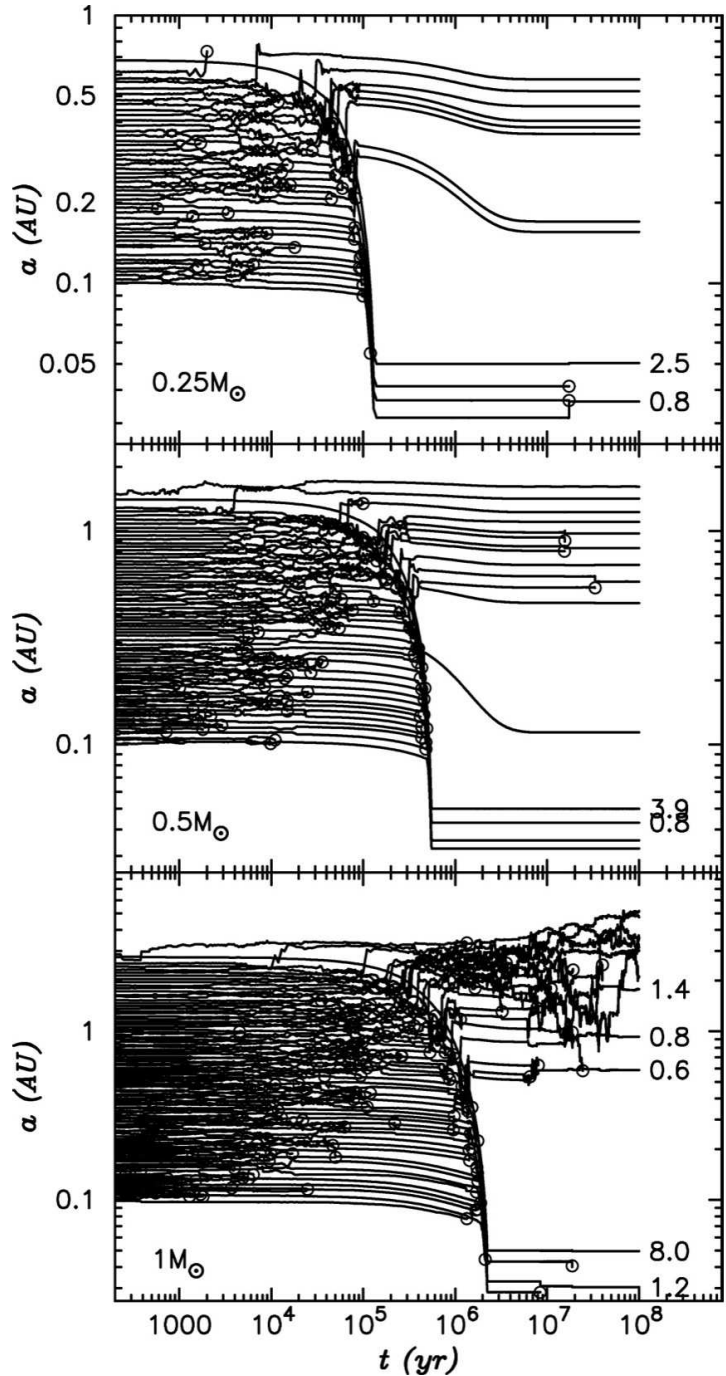}}}%
\caption{Migration of planetary embryos around a (from top to bottom) 
0.25, 0.5, and 1 solar-mass star. The outer planets were initially 2, 3.2, 
and 6 Earth-masses, respectively.
As the embryos migrate, they collide and grow to larger objects.
The numbers next to the lines in each panel correspond to the final mass 
of that body in Earth-masses. Figure from \cite{Kennedy08b} with the permission of AAS.}%
\label{Kennedy08}
\end{center}
\end{figure}

\vskip 20pt
\subsubsection{Formation of Super-Earths Around Low-Mass Stars: Disk Instability}

As explained before, given the low masses of the circumstellar disks around M stars, 
the existence of giant planets around these stars suggests that they might have formed at large 
distances and migrated to their current orbits. 
This assumption is based on the fact that in a planet-forming nebula, more material
is available at outer regions which can then facilitate the formation of a
giant planet through the core-accretion model. The availability of more mass at the 
outer distances in a disk prompted researchers to look into the possibility of
explaining the formation of super-Earths around M stars through the disk instability model.
Recall that in this scenario,
clumps, formed in an unstable gaseous disk, collapse and form gas-giant
planets (e.g. \cite{Boss00a,Mayer02}). 
After the giant planets are formed,
a secondary process is needed to remove their gaseous envelopes.
Simulations by Boss (2006; \cite{Boss06}) have shown that such collapsing clumps
can form around a 0.5 solar-mass star at a distance of approximately 
8 AU. This author suggests that, as most stars are formed in clusters
and in high-mass star forming regions, intense FUV/EUV radiations from
near-by O stars may rapidly (within 1 Myr) photo-evaporate the 
gaseous envelope around giant planets, leaving them with large 
super-Earth cores (Figure 17). Similar mechanism has been suggested for the formation
of Uranus and Neptune in our solar system \cite{Boss02}. A subsequent migration, similar to 
that suggested by Michael et al (2011, \cite{Michael11}), may then move these cores to close-in 
orbits.

The above-mentioned combination of the disk-instability and gas photo-evaporation 
presents a possible scenario for the formation of super-Earths 
at large distances and their migration to their closer orbits. However, this mechanism
does not seem to be able to explain the formation of the 
close-in 7.5 Earth-masses planet of the M star GJ 876 and its current 2-day orbit. 
According to the disk instability model, this object has to have
1) formed at a large distance where it also developed a gaseous envelop,
2) migrated inwards while its atmosphere was photo-evaporated, and finally
3) switched orbits with the two giant planets of the system--a scenario that (without
switching orbits) may be applicable to the formation of the recently detected outermost 
super-Earth planet of this system \cite{Rivera10}, but is very unlikely to have happened 
to the innermost body.

\begin{figure}
\begin{center}
\center{
\resizebox*{11cm}{!}{\includegraphics{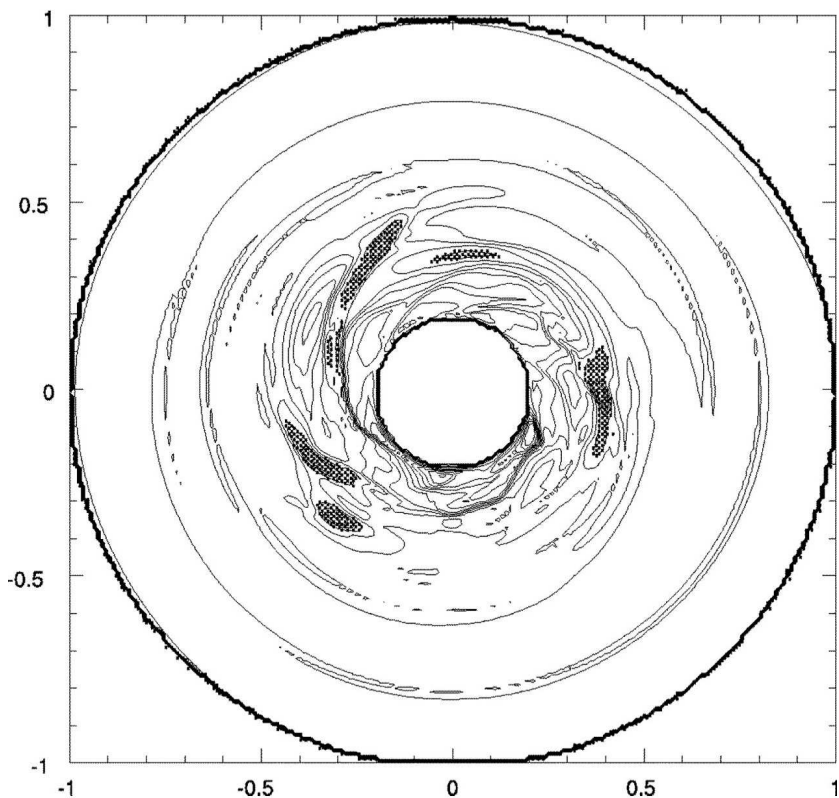}}}%
\caption{Formation of clumps (shaded regions) in a gravitationally unstable disk
around a 0.5 solar-mass star, after 215 years. The outer edge of the disk is 
at 20 AU and its inner radius is 4 AU. The clumps' densities are larger than
$10^{-10}$ g/cm$^{-3}$. Figure from \cite{Boss06} with the permission of AAS.}%
\label{Boss-Mstar}
\end{center}
\end{figure}

\vskip 20pt

As evident from the review presented in this section,
it is generally accepted that super-Earths are formed through 
a combination of a core accumulation process and planetary migration. 
Modeling the formation of these objects requires the 
simulation of the collisional growth of planetary embryos, and their
subsequent interactions with the protoplanetary disk. 
A realistic model requires a global treatment of the disk and inclusion
of a large number of planetesimals and planetary embryos.
In practice, such simulations are 
computationally expensive. To avoid such complications, most of 
the current models of super-Earth formation include only small 
numbers of objects (cores, progenitors, protoplanets, planetesimals, etc.). 
Recently McNeil \& Nelson (2010, \cite{McNeil10}) have shown that in systems with a large 
number of bodies (e.g. several thousand planetesimals and larger objects), 
the combination of the traditional core-accretion and
type-I planet migration may not produce objects larger 
than 3-4 Earth-masses in close-in (e.g., $\leq 0.5$ AU) orbits. 
Although the systems studies by these authors carry some simplifying 
assumption, their results point to an interesting conclusion:
while the combination of core-accretion and planet migration 
seems to be a viable mechanism for the formation of close-in super-Earths, 
the formation of these objects is still an open question, and a comprehensive 
theory for their formation requires more sophisticated computational modeling.

\section{Habitability of Super-Earths}

An important characteristic of super-Earths that differentiates them from
other planetary bodies (i.e., terrestrial and giant planets) is 
the masses of these objects. The larger-than-terrestrial masses of these planets
imply that super-Earths have the capability of developing and retaining 
atmospheres, and may also be able to have a dynamic interior. 
As super-Earths are formed (or dynamically evolved) in a region of 
a protoplanetary disk where the gas has a short lifetime, the amount
of the gas accreted by these objects, or trapped in their interiors when
they were fully formed, is much smaller than those of gas-giant planets. 
It is therefore natural to expect super-Earths to have thin to moderately
thick atmospheres (e.g., see \cite{Miller10} and \cite{Bean10} for developments
on modeling the atmosphere of super-Earth GJ 1214b, and its observational
constraints). Around small and cool stars such as M dwarfs, where the liquid water 
habitable zone is at close distances,
the thin to moderate atmospheres of close-in super-Earths and their probable
dynamic interior make these objects prime candidates for habitability. 
Such close-in habitable super-Earths are potentially 
detectable by both the ground- and space-based telescopes.
In this section, these unique characteristics of super-Earths are discussed 
in more detail.

It is important to note that 
the notion of habitability is defined based on the life as we know it.
Since Earth is the only habitable planet known to humankind, the orbital
and physical characteristics of Earth are used to define a
habitable planet. In other words, habitability is the characteristic
of an environment which has similar properties as those of Earth,
and the capability of developing and sustaining Earthly life. 

The statement above implies that the fact that the only habitable 
planet we know is Earth has strongly
biased our understanding of the conditions required for life. From
the astronomers' point of view, and owing to the essential role that
water plays on life on Earth, the definition of a habitable
planet is tied to the presence of liquid water. However as simple this
definition might be, it has strong connections 
to a variety of complex interdependent
processes that need to be unraveled and understood to make predictions
on which planets could be habitable.
The basic principle is that the surface temperature and pressure 
of a planet should allow for liquid water. This is determined by the 
amount of irradiation that the planet receives 
from the star, and the response of the planet's 
atmosphere. The latter delicately
depends on the composition of the planet, and that in turn determines 
the heat transport  mechanism, cloud presence, and many other atmospheric
properties.

The irradiation from the star is contingent on the type of the star
and the planet's orbital parameters. The atmospheric composition, on the other
hand, depends on the in-gassing, out-gassing, and escape histories
of the planet. The in-gassing and out-gassing accounts are intrinsically
connected to the interior dynamics of the planet, while atmospheric
escape is related to a variety of thermal and non-thermal processes,
which themselves are linked to the presence of a magnetic field. 
It is not clear how delicate the balance between these different processes
could be. Nor is it evident if there are different pathways that could yield a habitable
planet. However, the fact that Earth has succeeded in developing life indicates that
our planet might have followed
one, perhaps of many evolutionary paths that resulted naturally in
a complex system by the series of steps and bifurcations that it encountered.
It is important to note that
the complexity and interdependence of these processes cannot be taken as evidence
for the uniqueness of life on Earth. 
The road ahead is to understand which planetary characteristics 
are indispensable, which are facilitating, and which are a byproduct of evolution.
For that purpose, and in
order to assess the possibility that a planet (e.g., a super-Earth)
may be habitable, a deep understanding
of these processes (i.e. interior composition and dynamics, planet's 
magnetic field, and atmospheric characteristics) is required.

\subsection{Interior}

There are at least three aspects of the interior of a super-Earth that relate
to its possible habitability: its composition, the manifestation of plate tectonics,
and the presence of a magnetic field.

\begin{figure}
\begin{center}
\center{
\resizebox*{12cm}{!}{\includegraphics{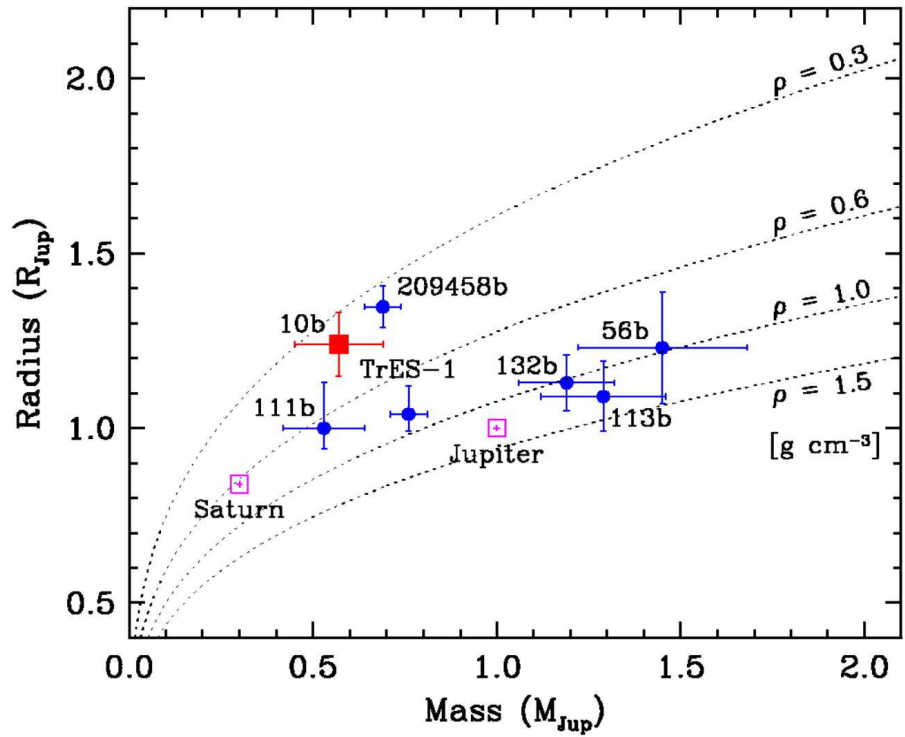}}}%
\caption{Graph of the mass-radius relationship for extrasolar planets OGLE-TR-10b 
\cite{Konacki05}, HD 209458b \cite{Brown01},
OGLE-TR-56b \cite{Torres04}, OGLE-TR-132b \cite{Moutou04}, OGLE-TR-111b \cite{Pont04}, 
OGLE-TR-113b \cite{Bouchy04,Konacki04},
TrES-1 \cite{Sozzetti04}.
The dotted lines represent curves of constant densities. Jupiter and Saturn
are also shown. Figure from \cite{Konacki05} with the permission of AAS.}%
\label{Mass-Radius}
\end{center}
\end{figure}

\subsubsection{Composition}

As water is the most essential element for habitability, one might expect that
it will have a significant contribution to the total mass of a habitable planet. 
However, on Earth, water constitutes less than 0.1\% of Earth's mass which places
Earth among the rocky planets of our solar system. This suggests, in order to study the 
habitability of extrasolar planets, it is important to distinguish planetary type, 
and identify rocky planets with some liquid water.

Unfortunately at the moment, the type of data available to infer the composition
of extrasolar planets is limited. The first generation of data comprises masses and radii
obtained from radial velocity and transit photometry searches. Neither of these quantities
alone can lead to a definitive  determination of the composition of a planet. However, 
for those planets whose masses and radii are known, a relationship 
between these quantities (known as the {\it mass-radius relationship})
can help to gain an insight on the possible materials that contributed 
to the formation of these objects.

Among the currently known extrasolar planets, the knowledge of both mass and radius 
is limited to only a small number of these bodies. The majority of these planets are
Jupiter-like with masses larger than 0.3 Jupiter-masses.
To the zeroth order of approximation, one can assume that these planets are 
perfectly spherical and have uniform interiors\footnote{The mass and radius of a spherical
body with uniform density $\rho$ are related as $M= (4\pi/3)\rho {R^3}$, where $R$ is the
radius and $M$ is the mass of the object.}. The mass-radius relationship in this case
will be of the simple form $R \sim {M^{1/3}}$. 
Figure 18 shows a few of these extrasolar planets in a mass-radius diagram [see also figure 4 of
Seager et al (2007, \cite{Seager07})].
The masses of these objects are in the range of 0.5 to 1.5 Jupiter-masses.
The figure shows some curves of constant density as well.
As shown here, the assumption of a uniform interior places these objects
close to the curves of constant density ranging from  
0.4 to 1.3 g cm$^{-3}$ \cite{Konacki05}.
As a point of comparison, Jupiter and Saturn are also shown. For more details,
we refer the reader to the paper by Seager et al (2007, \cite{Seager07}) on
the mass-radius relationship in massive extrasolar planets, and to the article by 
Sotin et al (2007, \cite{Sotin07})
where these authors discuss the mass-radius relationship of ocean planets.

\begin{figure}
\vskip 0.1in
\begin{center}
\center{
\resizebox*{15cm}{!}{\includegraphics{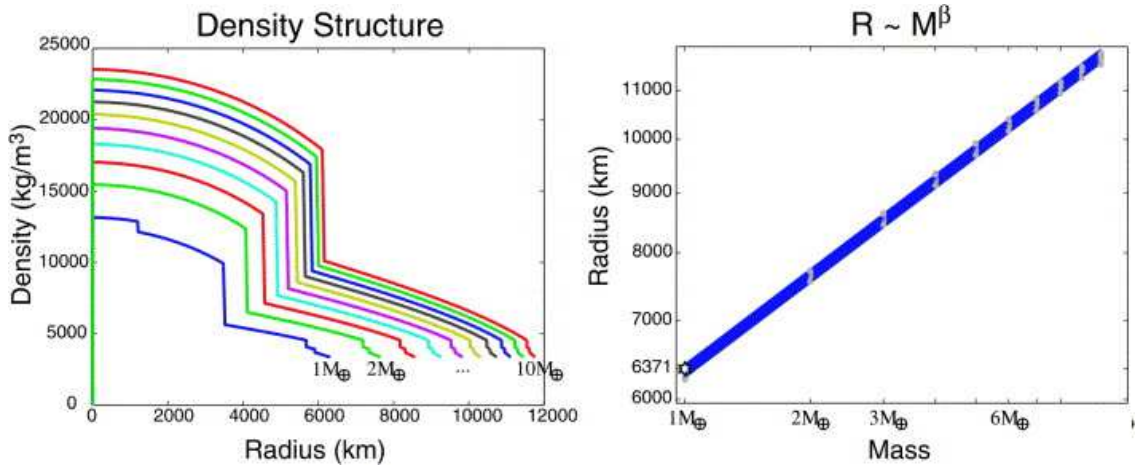}}}%
\caption{Left: Graph of the density of planets with masses ranging from 1 to 10 Earth-masses. 
The core of each planet constitutes 32.59\% of its mass. Each planet has 10\% iron in its mantle
and 30\% Ferromagnesiowustite in its lower mantle. Right: Mass-radius power law for planets
with ten different mineral compositions. The star denotes Earth. Note that the $x$-axis is
logarithmic. See \cite{Valencia06} for more details. Figure from \cite{Valencia07a} 
with permission.}%
\label{Valencia06}
\end{center}
\end{figure}

Despite the  apparent agreement between the values of the densities of the giant 
planets in Figure 18 and the assumption of a uniform interior, this assumption is
not valid for super-Earth objects. 
As known from Earth, the large amount of internal pressure in terrestrial
planets\footnote{In case of super-Earths, this pressure may amount to 
approximately 60 Mbars} causes the interiors of these objects to not have 
a uniform composition.
As a result, the mass-radius relationship for these planets deviates 
from the 1/3 power-law. 
Valencia et al (2006, 2007; \cite{Valencia06,Valencia07a})
studied these deviations for objects with masses of 1 to 10 Earth-masses.
Scaling Earth to larger sizes and assuming a layered structure
with different values of density, temperature, and pressure for each layer,
these authors modeled the composition of super-Earths by integrating the equation 
of state of each layer for different combinations of components
such as iron, silicate, magnesium, alloys, and water. The results of their simulations 
show that super-Earths may be mainly composed of iron cores, silicate mantles,
and water/ices (H$_{2}$O and ammonia, methane in minor proportions).
These authors suggested that the mass-radius relationship for super-Earths
may be of the form $R\sim A(R)\,M^{\beta}$ where the coefficient $A(R)$ has 
different values for different compositions,
and the exponent $\beta$ varies in a
small range between 0.262 and 0.274 (Figure 19). 

Although the results of the simulations by Valencia et al 
(2006, 2007; \cite{Valencia06,Valencia07a}) portray a general
picture of the components of which super-Earths might have formed, 
the mere knowledge of the mass and radius of
these objects is not sufficient to determine their actual compositions.
Since the above-mentioned mass-ratio relation is model-dependent, many combinations of
different components may result in the same mass and radius.
In other words, the mass-radius relationship suffers from a degeneracy that
stems from the existing trade-offs between components with different densities 
(iron cores, silicate mantles, water/icy layers, hydrogen envelope) \cite{Adams08}.
This degeneracy does not allow for the definite determination of the composition
of super-Earths.  

\begin{figure}
\vskip 0.1in
\begin{center}
\center{
\resizebox*{9.5cm}{!}{\includegraphics{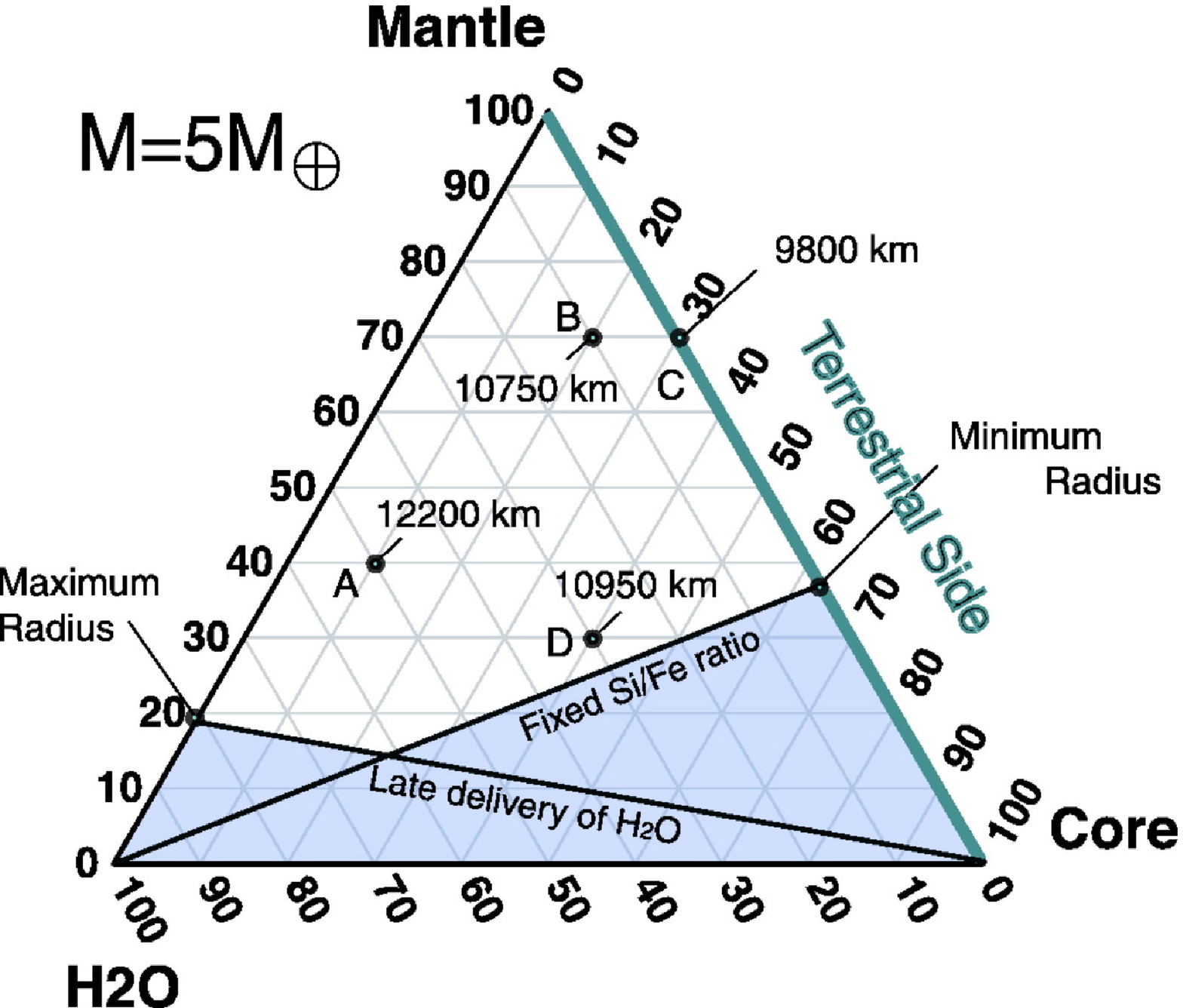}}
\vskip 20pt
\resizebox*{10.5cm}{!}{\includegraphics{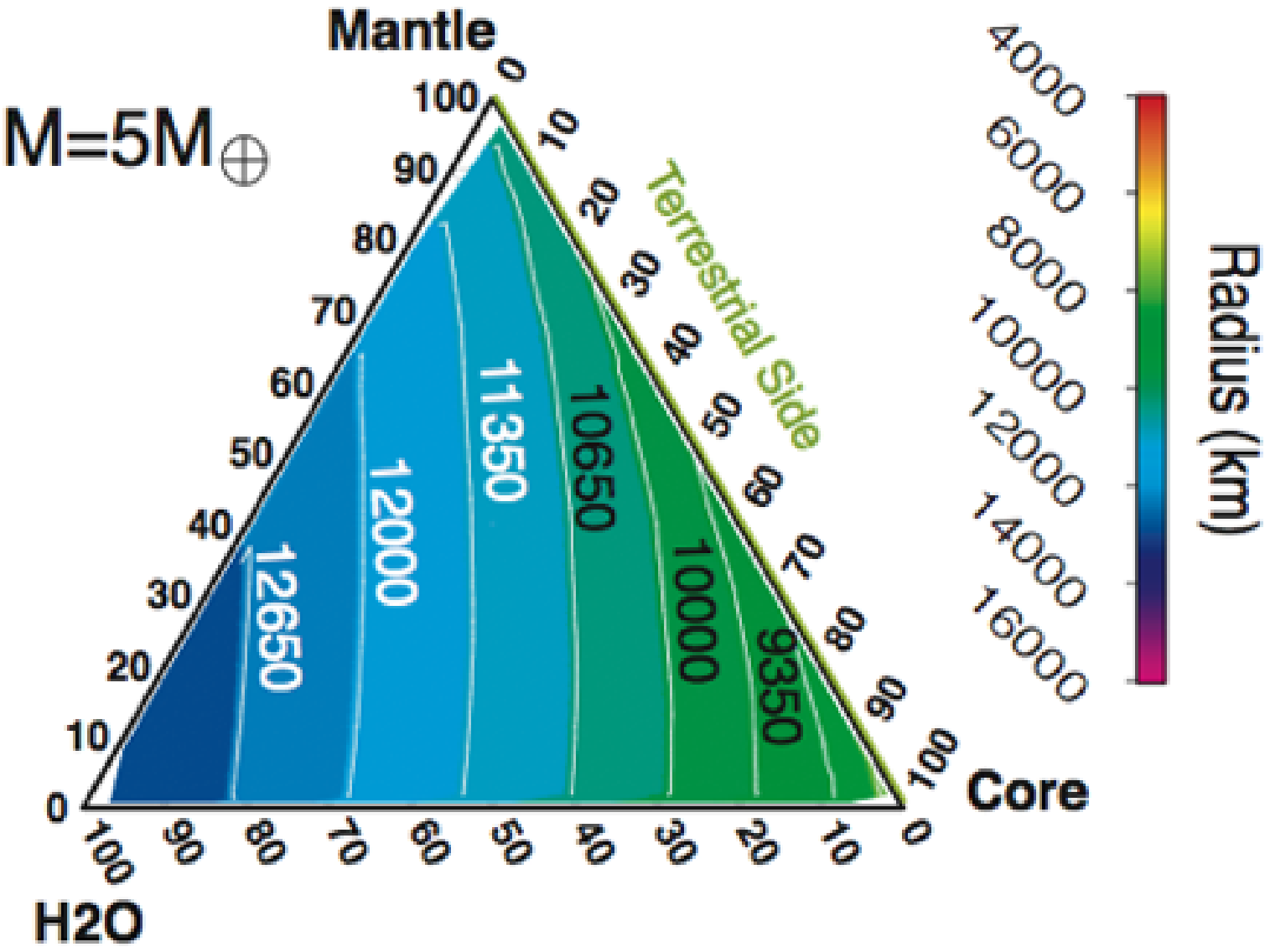}}}%
\caption{Ternary Diagrams for a 5 Earth-mass planet (e.g., GL 581 c) 
showing its possible compositions and different radii. 
Each diagram has three axes indicating the amount of water, iron
(core), and silicate (mantle). Each point inside a diagram 
corresponds to a unique combination of the three end-members. 
The vertices correspond to a 100\% composition, and the opposite 
line corresponds to 0\% of a particular component. A few examples
are shown in the top panel. Many different combinations of the three 
end-member components for a given mass can have the same radius. 
The bottom panel shows this in more detail. Labels on isoradius
curves are radii in km. A ternary diagram exists for each value of 
planetary mass.  The color bar 
spans the range of sizes of rocky and icy super-Earths  from 1 
Earth-mass pure iron planet (of 5000 km), to a 10 Earth-masses snowball planet 
(16000 km). Figure from \cite{Valencia07b} with the permission of AAS.}%
\label{Ternary}
\end{center}
\end{figure}

Despite this degeneracy, it may still be possible to attribute
a set of probable compositions to a super-Earth once its
radius is determined from observation. 
Integrating the equation of state for different combinations 
of silicate, iron, and water, and for different values of the radius of
a super-Earth with a known mass, Valencia et al (2007; \cite{Valencia07b})
have developed an archive that can be used for this purpose.
Figure 20 shows the results of one of their simulations.
Known as a {\it ternary diagram}, each panel of this figure shows 
the connection between different combinations of a 5 Earth-mass super-Earth
and its radius. Each vertex of a triangle represents an object with a 100\% 
composition of the vertex's material. Each side depicts the amount of the two
components on its two vertices that compose the planet.
A point inside the triangle uniquely specifies a
composition and its corresponding radius. As shown by the bottom panel,
super-Earths with similar masses but different compositions 
may have identical radii.

\begin{figure}
\begin{center}
\center{
\resizebox*{9cm}{!}{\includegraphics{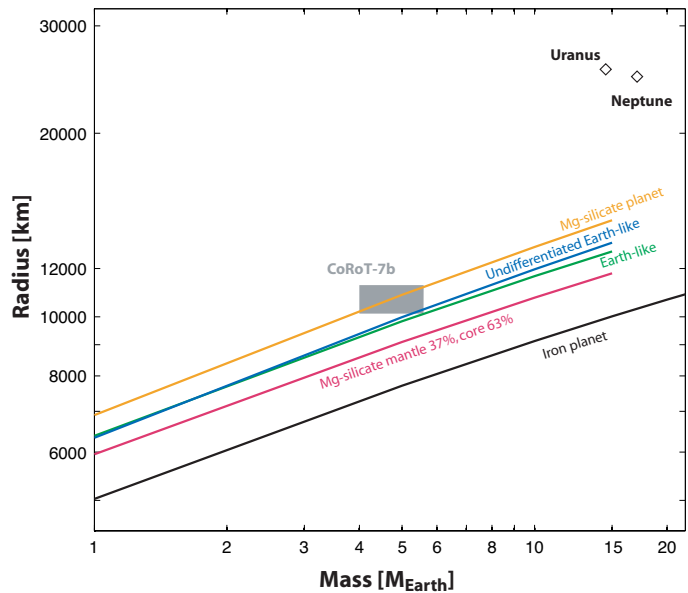}}}%
\caption{Mass-Radius relations for rocky super-Earths.  The shaded area shows 
the data for CoRoT-7b.  Five different compositions are shown. Orange: 
a Mg-Si planet (or super-Moon planet) has little or no content of 
iron;  Green: an Earth-like composition (differentiated planet with 
33\% core, 67\% mantle, and 0.1 molar iron fraction in the mantle); Blue: 
undifferentiated planet with the same elemental bulk composition as Earth 
(Fe/Si = 2);  Pink: A super-Mercury composition; Black: a pure iron 
composition.  The pure Mg-Si  and pure Fe compositions correspond to 
the lightest and densest rocky compositions, although they are highly 
unlikely given that Mg, Si, and Fe condense at similar temperatures.  
Compositions with increasing amounts of iron lie progressively below 
the Mg-Si planet composition. Any mass-radius combination that lies 
above the Mg-Si planet line necessarily implies a volatile content.
Figure from \cite{Valencia10} with permission from A\&A.}%
\label{CoRoT7b}
\end{center}
\end{figure}

Using a ternary diagram, it is possible to identify the extreme sizes
that a planet might have. For instance, from Figure 20,
the maximum value of the radius of a 5 Earth-masses super-Earth corresponds 
to a planet that 
is formed entirely of pure ice and water (left corner of the ternary diagram).
A radius larger than that of such a snowball planet would indicate the presence
of an atmosphere which could probably be made of hydrogen/helium. 
The minimum value of the radius of a super-Earth, on the other hand, 
corresponds to a planet that is 
made of pure iron or heavy alloys (right corner of the ternary diagram).
There is also a maximum size for a rocky planet corresponding to a pure silicate 
composition devoid of an iron core. Any radius above this critical size would indicate
the presence of volatiles. By volatiles we refer to water and other ices 
(ammonia, methane), as well as hydrogen and helium. The progression of the radius
from the dry side (mantle-core connection) to the wet side suggests that
for a given planet, there is a threshold radius beyond which the planet contains
a substantial amount of water (e.g., an ocean planet). This threshold corresponds
to the largest isoradius curve that intersects the terrestrial side of the
ternary diagram. For a 5 Earth-mass super-Earth, as shown in the lower panel 
of Figure 20, this radius is equal to 10400 km (not shown in the 
figure) \cite{Valencia07b}.  

As mentioned earlier, in order to obtain an insight into possible scenarios for the
composition of a super-Earth, the values of its mass and radius have to be known. 
Among the currently known super-Earths, CoRoT-7b \cite{Leger09} and 
GJ 1214b \cite{Charbonneau09} are two planets
for which these values have been determined. The knowledge of the
orbital elements and mass-radius of these planets has made it possible to obtain 
a better understanding of the compositions of these bodies. 
For instance, as shown by Valencia et al (2010; \cite{Valencia10}),
and following the numerical modeling of 
Valencia et al (2006, 2007; \cite{Valencia06,Valencia07a}),
the best fit to the size and mass of CoRoT-7b points to a composition with 3\% water vapor 
above a rocky interior. Within a one-sigma uncertainty, the composition could range from at
most 10\% vapor to an Earth-like composition with 67\% silicate mantle and
33\% iron core (Figure 21, also see Swift et al 2010 \cite{Swift10}).
In addition, given its proximity to its Sun-like 
star, CoRoT-7b is highly irradiated. Such
strong irradiation causes significant atmospheric and mass loss \cite{Jackson10}.
Given that the evaporating flow of an exoplanet
has already been observed as in the case of the transiting planet
HD 209458b \cite{Vidal03}, in the future, it might
be possible to detect the nature of the evaporating flow of CoRoT-7b
as either silicate- or vapor-based, since the limiting factor is not
the size of the planet but the star brightness.

\begin{figure}
\begin{center}
\center{
\resizebox*{12cm}{!}{\includegraphics{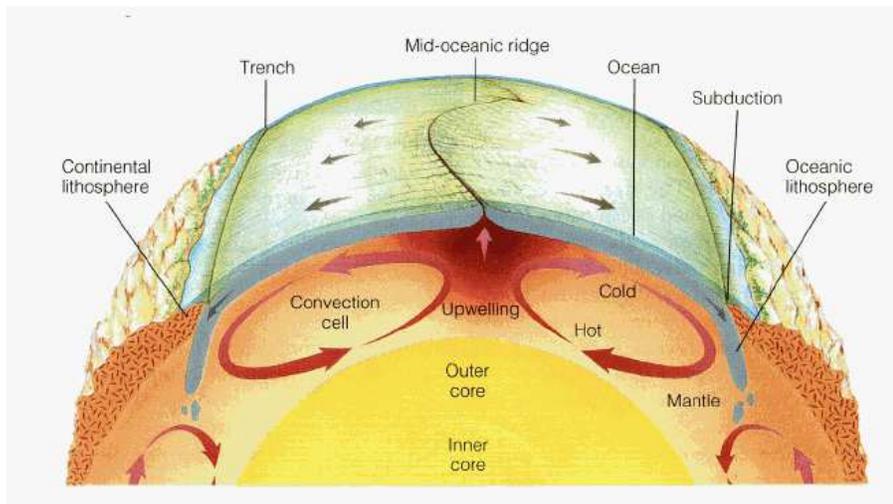}}}%
\caption{Artistic illustration of conduction cells in Earth's mantle.}%
\label{Mantle-Convection}
\end{center}
\end{figure}

\vskip 20pt
\subsubsection{Plate Tectonics}

Given the similarity between the formation of super-Earths and terrestrial planets, 
it is expected that these objects are formed hot, and have hot interiors. As in Earth,
the internal heat of these bodies may be due to the radioactivity of unstable elements, 
as well as processes such as the impact of planetesimals and planetary embryos during the
formation of these objects, their gravitational contraction during their accretional growth, 
and frictional heating due to the settling of heavy elements in their cores 
(the differentiation process).

Although the actual composition of super-Earths is unknown, those that are 
potentially habitable are expected to be mainly made of rocks. 
As the surface layers of these planets cool from above and form a crust, the heat generated by  
the above-mentioned processes will be trapped inside and produce
large convection cells within the mantles of these bodies\footnote{The internal heat of a planet 
is transported out through both conduction 
and convection. However, because rock is a poor conductor, the amount of the heat transported 
through conduction is negligibly small.}. These convection cells cycle hot material through the 
mantle and gradually cool the planet. The cooling of the mantle through convection also
controls the cooling of the core. Figure 22 shows an artistic conception of this process.

Convection may operate in two different modes: mobile and non-mobile. In the non-mobile
mode, the material cycled by convection cells forms a rigid and immobile layer at 
the surface of the mantle known as {\it stagnant-lid} \cite{Lenardic04,Loddoch06,Reese05}. 
During stagnant-lid convection, the surface plate thickens with time and acts as a lid to 
the interior. Compared to the mobile mode, the stagnant-lid regime is an 
inefficient mode of cooling a planet.
The interior heat in this mode may transfer out through volcanism as in the moon
of Jupiter, Io \cite{Loddoch06}, or may gradually be transported by the lid through conduction. 
In the mobile mode, on the other hand,
the lithosphere or plate actively participates in convection by being formed at
mid-ocean ridges and subducted at trenches. Known as {\it plate tectonics}, 
this mechanism allows for a more efficient heat and chemical transport from the interior 
to the surface of the planet (Figure 22). The recently suggested remote detection of volcanism on 
exoplanets by Kaltenegger et al (2010, \cite{Kaltenegger10a}) may be useful in identifying
possible modes of cooling of a planet and to determine if the planet undergoes plate tectonics.

The mode of convection (stagnant-lid with volcanic activity like Io or without 
profuse volcanism vs. plate tectonics) has a profound effect on the thermal evolution
(and consequently the habitability) of a rocky planet. 
Among the terrestrial planets in our solar system, Earth is the only one
with an active plate tectonics. While Venus has a similar mass, its
heat has been transported out through a stagnant-lid process for at least
500 Myr. Mars, with its small size, also has stagnant-lid convection,
although it might have had plate tectonics sometime in the past \cite{Lenardic04}.

The reason for plate tectonics on Earth and not on other solar system objects is 
still under debate. It is widely accepted that this mechanism has played 
an important role in the geophysical evolution of our planet, and is associated
with its geochemical cycles. As a result, plate tectonics has been recognized
as an important mechanism for the habitability of Earth \cite{Walker81}. However,
whether this process exists (or should exist) in any habitable planet
is unknown. Although it seems natural to assume that, similar to Earth, any habitable planet
has to have a dynamic interior and maintain plate tectonics, it is not clear whether
that is entirely true. In regard to super-Earths, as explained below, the matter is 
even more complicated.

The subject of plate tectonics on rocky super-Earths is controversial.
Much research has been done on this topic and depending on the approach to
mantle convection, results point to two different schools of thought: 
favoring plate tectonics based on scaling mantle convection in Earth 
to larger planets, and favoring a stagnant-lid regime based on numerical modeling
of convection in the mantles of super-Earth objects.

On the scaling mantle convection, Valencia et al 
(2007, 2009; \cite{Valencia07c,Valencia09}) proposed that 
massive terrestrial planets would have more favorable conditions for subduction, which
is an essential part of plate tectonics. In their model, these authors used a parameterized
convection scheme described in terms of the surface heat flux, and
included the effects of compression in the structure parameters (mantle
thickness, gravity, etc). They concluded that while faults' strength
increases with mass, the convective stresses increase even more, so
that deformation can happen more easily in massive planets. This is
due to the canceling effect between increasing the gravity and decreasing the thickness
of the plate which causes the pressure-temperature regime of
the plate to be almost invariant with size.  
Valencia et al. (2007, 2009; \cite{Valencia07c,Valencia09}) also suggested that
unlike small planets such as Earth, where plate tectonics would depend on the 
presence of water, larger terrestrial planets have sufficiently large 
convective stresses and would not need weakening agents 
to lower their yield stress in deformation. This implies that the one Earth-mass 
regime seems to be the lower threshold for active-lid tectonics. 

Another approach to plate tectonics in super-Earths is 
given by O'Neill \& Lenardic (2007; \cite{ONeil07}).
These authors  suggested that at most, massive Earth-analogs would be in an episodic
regime in which episodes of plate tectonics and stagnant-lid occur
at different times. They assumed that planets are in
a mixed heated state with different proportions of radioactive to
basal heating for each planet, and adapted the numerical model of 
Moresi \& Solomatov (1998; \cite{Moresi98}), which has been developed to reproduce 
plate tectonics on Earth, to model mantle convection in super-Earths.
Despite that Valencia et al (2007, 2009; \cite{Valencia07c,Valencia09}) and
O'Neill \& Lenardic (2007; \cite{ONeil07}) agree on considering the Byerlee 
criterion (an empirical relation to determine the minimum amount of stress 
that is required to fracture a planet's crust along its faults, \cite{Byerlee78}) 
for plate boundary creation and the need for convection-induced stresses, results by
O'Neill \& Lenardic (2007; \cite{ONeil07}) conclude that owing to higher
gravity, faults are locked due to increased pressure and thus deformation
is halted. 

Other recent studies on this topic have arrived at different results.
For instance, Tackley \& van Heck (2009; \cite{Tackley09}) used numerical
modeling and constant density scaling, and show that planets that are internally
heated as well as those heated from below (and maintain a temperature difference between 
top and bottom), are more likely to have plate tectonics
as in Earth. 
Sotin \& Schubert (2009; \cite{Sotin09}) have also attempted to explain the 
difference between the results obtained by  
Valencia et al. (2007, 2009; \cite{Valencia07c,Valencia09}) and 
O'Neill \& Lenardic (2007; \cite{ONeil07}). Utilizing a parameterized convection
approach and using the results of their structure-scaling model, these authors
have shown that despite an overestimate of the ratio of the driving to
resistive forces in the model by Valencia et al (2007; \cite{Valencia07c}),
this ratio is weakly dependent on the size of a terrestrial planet, and other
compositional and/or geophysical properties may have to be considered in
order to determine  the
probability of the occurrence of plate tectonics in super-Earths.
Sotin \& Schubert (2009; \cite{Sotin09}) also considered a 3D spherical
scaling and assumed an increase in the heat flux of a planet with increasing
its size, and showed that planets such as super-Earths may be marginally in
the plate-tectonic regime.

In conclusion, whether plate tectonics occur in super-Earths or not
is still under debate. Although there seems to be better qualitative agreements between 
models, there are still discrepancies that have to be resolved.

\vskip 20pt
\subsubsection{Magnetic Fields}

One important characteristic of Earth, that is a consequence of having a molten dynamic iron core
and an active and on-going plate tectonics, is its magnetic field. Earth's magnetic field plays 
an important role in its habitability. It protects our planet from harmful radiations and 
maintains its atmospheric composition by preventing non-thermal escape of 
different elements and components \cite{Griessmeier04,Lundkin07,Dehant07}. 
As such, the presence of a magnetic field has been considered
essential for habitability. 

Whether and how magnetic fields are developed around super-Earths is an active topic of research. 
In general, in order for a magnetic field to be in place around an Earth-like planet,
a dynamo action has to exist in the planet's core. In order for this dynamo to develop and
sustain, the planet has to have a core of liquid iron (or an alloy \cite{Stevenson10}) with a
vigorous and on-going convection process. 
The latter can be sustained by maintaining a temperature difference 
across core-mantle boundary which itself depends on the efficiency of transporting heat
and cooling the planet. On Earth, the core has cooled enough as to yield 
a freezing inner core which releases latent heat into the liquid outer core 
that drives Earth's dynamo. Also, thanks to plate tectonics,
the mantle is cooling effectively to allow for the core to sustain it super-adiabaticity 
(the hotter part of the core becomes less dense and rises to the cooler part in a fast pace).
We refer the reader to Planetary Magnetism, a special issue of Space Science Review
by Christensen et al \cite{Christensen10} for a complete review of the current state of research
on planetary magnetism.

The appearance of a magnetic field around a super-Earth and its lifetime are different
from those of Earth.  
Studies of the internal heating and cooling of these objects suggest that
large super-Earths will not be able to develop magnetic fields.
Modeling the internal evolution of hot super-Earths (i.e. super-Earths in close-in
orbits) and studying their cooling histories, Tachinami et al (2009; \cite{Tachinami09}) 
have shown that planets more massive than 5 Earth-masses would not be able to develop a 
dynamo for most of their evolution\footnote{The melting curve of iron shows a steep 
increases with pressure. s a result, in order for large planets to have a molten core, 
the temperature in the core has to rise to values of the order $10^4$ K. These
values are too high to be reached}. Recent study by Gaidos et al \cite{Gaidos10}
lowers this limit to 2 Earth-masses. As shown by these authors, planets larger than
1.5-2 Earth-masses with stagnant lids do not generate a dynamo. Only if in these planets,
the cooling of the core is supported by a mobile lid, they can produce magnetic 
fields that may last a long time. Figure 23 shows the results of some the simulations
by these authors. As shown here, CoRoT-7b might have maintained a magnetic field for the duration
of its lifetime.

Stamenkovic et al (2010; \cite{Stamenkovic10a}) and 
Stamenkovic \& Breuer (2010; \cite{Stamenkovic10b}) have also indicated that the 
possibility of developing a magnetic field decreases as the planet's mass becomes larger.
These authors studied the thermal evolution of planetary bodies with masses ranging from
0.1 to 10 Earth-masses, and showed that when a pressure-dependent viscosity is included
in their models, results suggest that mantle convection and the growth of
a low-lid in the core-mantle boundary will be ineffective. According to these authors, 
the heat-transport through convection
will eventually cease and the cooling of the core will be only through conduction. Since
conduction is not an effective way to transport heat from the core, the thermally generated
magnetic field will be strongly suppressed. The results of the simulations by 
Stamenkovic et al (2010; \cite{Stamenkovic10a}) and 
Stamenkovic \& Breuer (2010; \cite{Stamenkovic10b}) also
suggest that the scaling laws, as used by Valencia et al (2007; \cite{Valencia07c}) and
O'Neill \& Lenardic (2007; \cite{ONeil07}), cannot be used for pressure-dependent viscosity 
models such as those for studying the interior of super-Earths.

\begin{figure}
\vskip 0.1in
\begin{center}
\center{
\resizebox*{12cm}{!}{\includegraphics{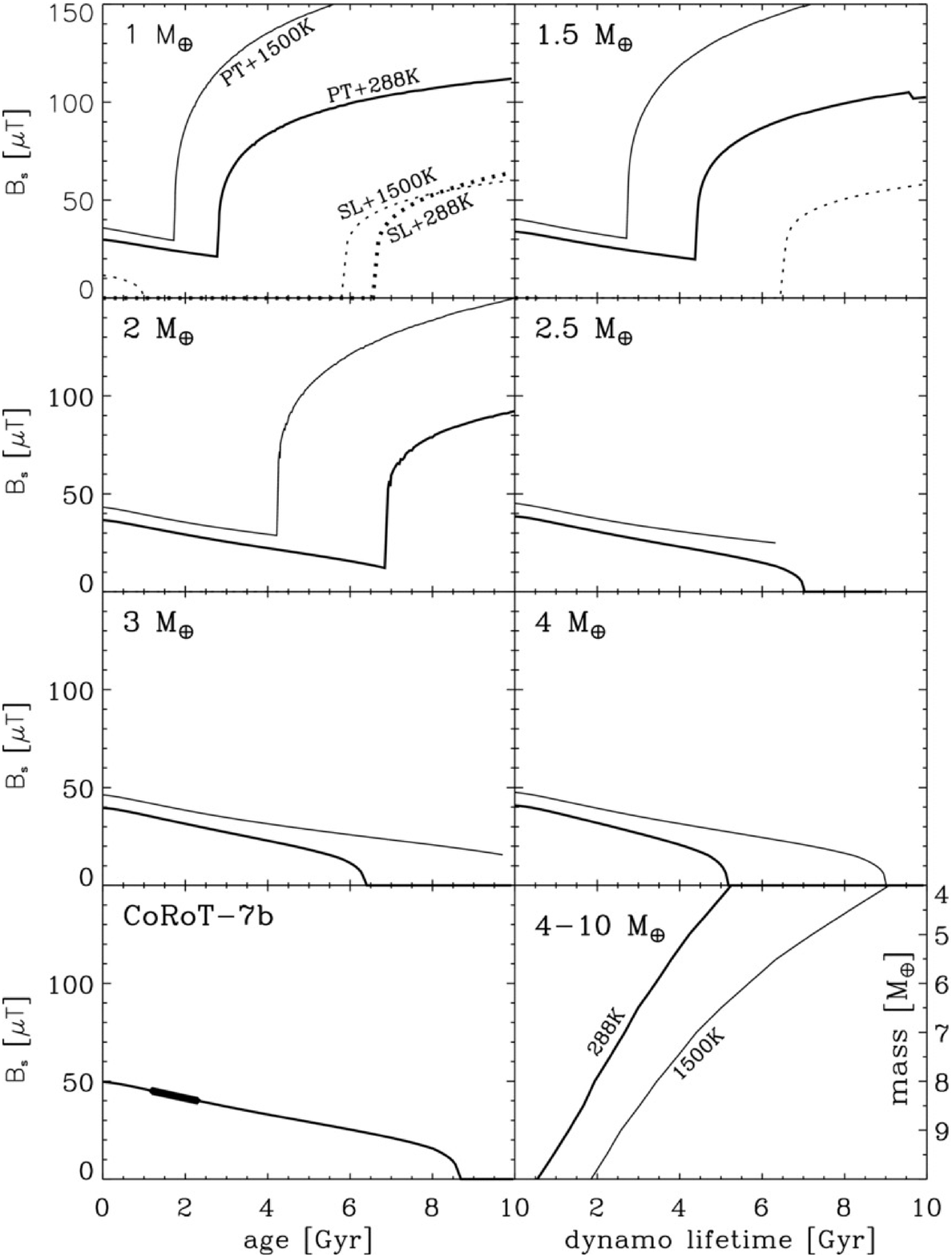}}}%
\caption{Averaged surface magnetic fields of super-Earths. PT and SL in the graph
of a 1 Earth-mass planet stand for plate tectonics and stagnant lid. The temperature
of each graph corresponds to the effective surface temperature of the planet. As shown here,
planets larger than 2.5 Earth-masses with stagnant lids do not develop magnetic fields.
For those with plate tectonics, the lifetime of the magnetic field decrease as the
mass of the planet increases. Note that for simulations of CoRoT-7b, the surface temperature 
of the planet was set to 1810 K \cite{Leger09}. Figure from \cite{Gaidos10} 
with the permission of AAS.}%
\label{Magnetic-Field}
\end{center}
\end{figure}

\subsection{Atmosphere}

The presence of an atmosphere around a terrestrial planet has profound effects on its capability 
in developing and maintaining life. While the chemical properties of the atmosphere point to the
planet's possible biosignatures \cite{Kaltenegger07,Kaltenegger09,Kaltenegger10b} as well as the 
materials of which the planet is formed (the latter can be used to infer information about the
origin of the planet, its formation mechanism, as well as its orbital evolution and interior 
dynamics), 
its greenhouse effect prevents the planet from rapid cooling,
and its cloud circulations enable the planet to maintain global uniformity
in its surface temperature. As such, the planet's atmosphere plays an important role in
the determination of the habitable zone of its central star.

\begin{figure}
\begin{center}
\center{
\resizebox*{13cm}{!}{\includegraphics{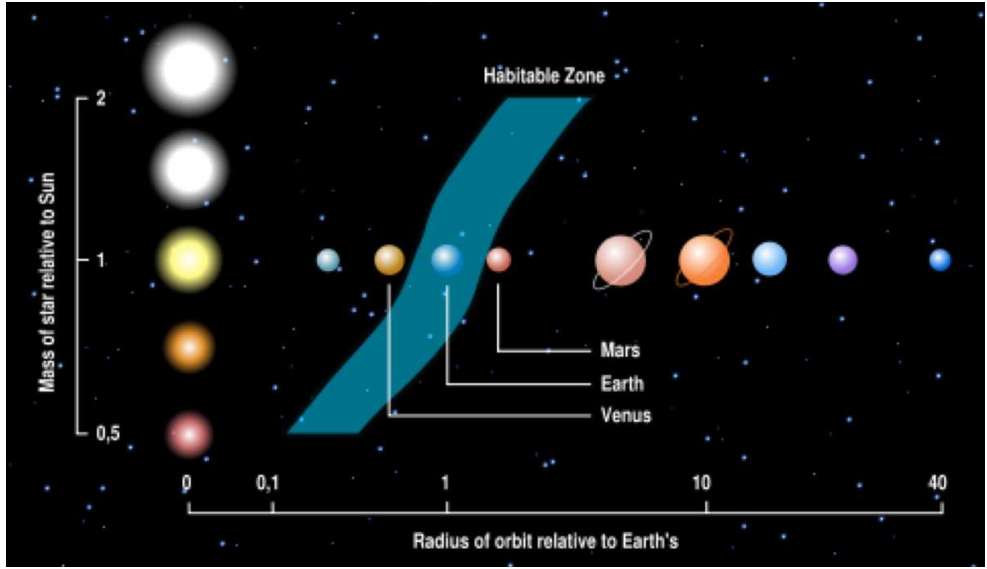}}}%
\caption{Habitable zones of stars with different masses based on the model 
by Kasting et al (1993; \cite{Kasting93}).}%
\label{HZ}
\end{center}
\end{figure}

In general, the habitable zone of a star is defined as a region where an
Earth-sized planet can maintain liquid water on its surface \cite{Kasting93} (Figure 24).
In the absence of planetary atmosphere, the width of this region is small and the locations
of its inner and outer boundaries are determined by the amount of the radiation that planet
receives from the central star. When the planet is surrounded by an atmosphere, the greenhouse 
effect causes these boundaries to move to larger distances. In this case, the outer edge of the 
habitable zone is defined as a distance beyond which ${\rm CO}_2$ clouds
can no longer keep the surface of the planet warm and runaway glaciation
may occur. Correspondingly, the inner edge of the habitable zone is defined as a distance closer
than which runaway greenhouse effect may increase the surface temperature and
pressure of the planet to values higher than those accommodating life
\cite{Forget97,Williams97,Mischna00}.

Whether or not a super-Earth can have an atmosphere, and what the chemical composition of
this atmosphere would be are directly linked to the properties of the environment 
where the super-Earth was formed, and it's subsequent interior dynamics and orbital evolution. 
As explained in section 2,
the fact that super-Earths are smaller than giant planets suggests that these 
objects might have either formed in the low-mass and gas-poor region of a 
protoplanetary disk, or were formed in its outer regions where the disk is more 
massive and the lifetime of the gas is longer, but were scattered to the 
inner orbits before they accreted a large amount of gas. As a result, one
can think of three mechanisms for the formation of an atmosphere around a
super-Earth: 

\begin{itemize}

\item direct accretion of the gas from circumstellar disk, 
\item out-gassing during the formation of the planet, and 
\item out-gassing due to the planet's active interior and plate tectonics.

\end{itemize}

Different mechanisms of the formation of
an atmosphere, combined with different scenarios for the formation of super-Earths lead to a range
of atmospheres with different masses and elemental-abundances.
For instance, as shown by Elkins-Tanton \& Seager (2008; \cite{Elkins-Tanton08}), a super-Earth
accreted from planetesimals of different primitive and differentiated
chondritic and achondritic meteorites can out-gas an atmosphere
with an initial mass ranging from approximately 1\% to a few percent of the total
mass of the planet (in some extreme cases the mass of the atmosphere may even
increase to over 20\% of the planet's total mass).
Considering an object with a mass of  1 to 30 Earth-masses, these authors have shown that
this atmosphere may be primarily made of water, hydrogen, and/or carbon compounds that include
oxygen. The abundance of helium in such an atmosphere will be very small, and its nitrogen concentration
will be equivalent to the amount of nitrogen in Earth's atmosphere.

\begin{figure}
\begin{center}
\center{
\resizebox*{11cm}{!}{\includegraphics{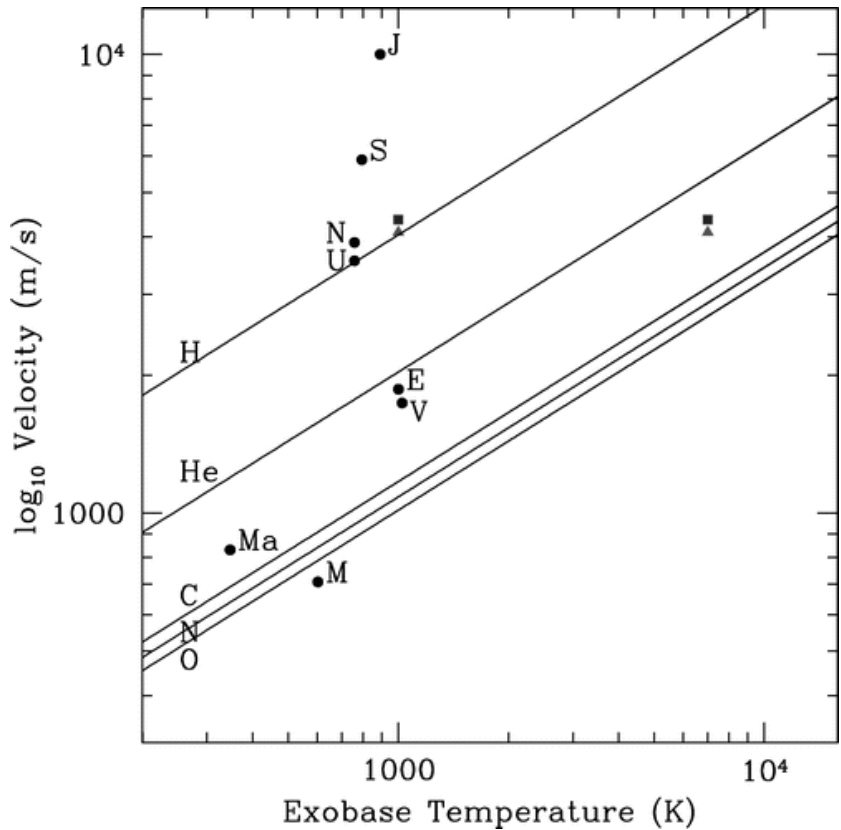}}}%
\caption{Graph of thermal velocity of different elements (solid lines) and escape
velocities of planets (shown by symbols) in terms of temperature. Planets of solar
system are shown by their initials. The triangles represent a 10 Earth-masses planet with 
similar differentiated composition as Earth, and the squares represent a 10 Earth-masses planet
with 25\% water in the outer layer, 52.5\% silicate mantle, and 22.5\%  iron core 
\cite{Seager07}. The exobase temperatures at 1000 K and 10,000 K correspond to those of super Earths 
in close-in orbits around M dwarfs and Sun-like stars, respectively. As shown here,
a planet will retain all gases that have thermal velocities below its escape velocity. 
Figure from \cite{Elkins-Tanton08} with the permission of AAS.}%
\label{HZ}
\end{center}
\end{figure}

Although the initial mass and elemental abundance of an out-gassed atmosphere
is driven by the material of which the planet is formed and by planet's geochemical and
geodynamical history, its final chemical composition is determined
by a series of processes such as thermal and non-thermal atmospheric escapes,
interactions of molecules with the light of the star (photolysis), and the
rates of such chemical interactions. Among these processes, atmospheric escape
plays a more important role. Molecules such as hydrogen, for instance, may undergo
hydrodynamics escape and carry off larger elements with them. This process may, however,
be counterbalanced by the surface gravity of the planet. The gravitational
attraction of a massive planet may prevent molecules from escaping its atmosphere,
or at least slow down the rate of their escape. In regard to super-Earths, the latter
implies that super-Earths more massive than Earth should be able to retain hydrogen in their
atmospheres and avoid its erosion through (thermal) atmospheric escape. Figure 25 shows
this by comparing a planet's escape velocity with the thermal velocity of different
elements in the models studied by Elkins-Tanton \& Seager (2008; \cite{Elkins-Tanton08}).
As shown here, a 10 Earth-mass planet in a short period orbit around a Sun-like star or
an M dwarf can retains its hydrogen and prevents it from thermal escaping.
Simulations by these authors also suggested that the initial amount of
hydrogen in the out-gassed atmosphere of a (massive) super-Earth may not exceed $\sim 6\%$
implying that if more hydrogen is detected
around a super-Earth, it must have come through other processes 
(e.g., as shown by Alibert et al \cite{Alibert06},
a 10 Earth-mass planet, similar to that in a 0.08 AU orbit around the star HD 69830,
may retain equivalent to 2 Earth-masses of H/He while migrating to a short-period orbit
and subject to atmospheric evaporation).

Other processes that erode an atmosphere include
intense radiation from the central star (e.g, extreme ultraviolet) \cite{Baraffe06}, stellar wind, 
and coronal mass ejection \cite{Lammer07}. 
Around M stars, in particular, these processes are strong, and compared to 
solar-type stars, they last longer \cite{Scalo07,West08}. As a result, many authors have 
assumed that close-in super-Earths around M stars may have small to no atmospheres 
\cite{Valencia06,Valencia07a,Valencia07b,Seager07,Sotin07,Fortney07,Lammer07,Selsis07,Seager09}.

Whether massive super-Earths, or those subject to small stellar radiations can have 
hydrogen-rich atmospheres, and how to identify the signature(s) of such an atmosphere 
have been subjects of active research for the past 
few years. The majority of these studies rely on the computational simulations of the accumulation
and erosion of a gaseous envelope around an Earth-sized or larger body
in order to estimate the amount of hydrogen that may exist
in the atmosphere of a super-Earth. These simulations themselves require
detailed modeling of processes such as atmospheric escape, photochemistry, and the 
pressure/temperature distribution in the lower part of planet's atmosphere (near the
surface of the planet). Since different models use different assumptions on the formation, evolution,
and interior dynamics of a super-Earth, the results of these simulations are different.
Some point to a hydrogen-rich atmosphere, whereas some suggest moderate 
to low levels of H/He, and some even present the possibility of other chemical compositions.  
An interesting case that has been studied by several researchers is the case of the transiting 
super-Earth GJ 1214b.
The larger-than-Earth radius of this planet (2.7 Earth-radii) combined with its low density 
$(\sim 1870\> {\rm kg \> m}^{-3})$ \cite{Charbonneau09} suggests that this planet has to be surrounded
by a gaseous envelope. As shown by Miller-Ricci \& Fortney (2010; \cite {Miller10}), a hydrogen-rich
atmosphere around this super-Earth may not be unrealistic, and can produce up to 0.3\%
variations (as a function of the wavelength) in the depth of the primary transit 
of this body compared to the background light of its parent M star. As shown by
Rogers \& Seager (2010a; \cite{Rogers10a}), on the other hand, a hydrogen-rich atmosphere
may not be the only possibility. Analyzing the bulk composition of GJ 1214b using the models
developed by the same authors \cite{Rogers10b}, Rogers \& Seager (2010a; \cite{Rogers10a})
considered three possible scenarios for the accumulations of an atmosphere around this planet;
accreting gas from the protoplanetary nebula, ice sublimation, and out-gassing.  
These authors have shown that if GJ 1214b accreted its gaseous envelope from the primordial
nebula, its interior would be primarily composed of iron, silicate, and ice, and its
gaseous envelope would contain primordial H/He with a mass equivalent to 0.01\% to 
5\% of the total mass of the planet. An atmosphere formed mainly from ice sublimation, 
on the other hand, would contain a massive amount of water, equal to at least 47\% of the 
planet's mass. Such an atmosphere would be able to account for the measured values of mass and 
radius of GJ 1214b without requiring a layer of H/He. Finally, an out-gassed
atmosphere would have to be rich in hydrogen in order to be able to account for the observed value
of the planet's radius. Observational uncertainties on the measurements of the radius of the
star GJ 1214, on the other hand, have put strong constraints on the predictions of these models.
As suggested by Charbonneau et al (2009, \cite{Charbonneau09}), 
the M star GJ 1214 may be 15\% smaller in radius which
suggest that the radius of GJ 1214b will be also smaller with the same amount, removing the 
necessity for a large atmosphere to explain its observed radius \cite{Charbonneau09}.

As shown in this section, the uncertainties in the internal composition of super-Earths 
result in different models for the atmospheres of these objects. 
A test of the validity of these models can come from the observation of the transmission and/or emission 
features in the spectra of these bodies.
As suggested by Miller-Ricci et al (2009; \cite{Miller09}), super-Earths with massive hydrogen atmospheres
will show strong water features in their emission spectra whereas those that lack hydrogen
show signatures of CO$_2$. Also, a hydrogen-dominated atmosphere will show significant 
spectral lines in its transmission spectrum since such an atmosphere has a large scale 
height\footnote{Transmission features in a planet's spectrum are direct indication of the scale height of
its atmosphere}. Recent observations of GJ 1214b have not been able to detect such transmission
features in the planet's spectrum implying that GJ 1214b does not have a cloud-free 
hydrogen-rich atmosphere. The lack of features in the transmission spectrum of this planet
seems to instead point to a dense steam atmosphere \cite{Bean10}.

In conclusion, it would not be unrealistic to assume that
super-Earths carry gaseous envelopes. Around low-mass stars, some of such atmosphere-bearing 
super-Earths may even have stable orbits in the habitable zones of their host stars. 
As we explain in the next section, many of such habitable super-Earths are potentially
detectable by different observational techniques. The recently detected super-Earth
GL 581 g \cite{Vogt10} with its possible atmospheric circulation  \cite{Heng10} in the habitable zone
of its host star may in fact be one of such planets. More advanced telescopes 
are needed to identify the biosignatures of these bodies and the physical and compositional 
characteristics of their atmospheres.

\section{Detection of Super-Earths}

\subsection {Radial Velocity Technique}

Although the current sensitivity of the radial velocity technique has reached a level that 
allows for the detection of super-Earths around solar-mass stars, low-mass stars
such as M dwarfs present the most promising targets for searching for Earth-sized planets and
super-Earths. This is primarily due to the fact that as the least massive stars, M dwarfs show 
the greatest reflex acceleration due to an orbiting planet. Also, simulations of planet formation
suggest that planets formed around M dwarfs are generally smaller than gas-giants and more probably 
in close-in orbits (note that Jovian-type planets have in fact been detected around M stars.
Also note that precision Doppler surveys are optimally sensitive to small orbits).
It is therefore not surprising that
during the past few years, 23 extrasolar planets have been detected around 
17 M stars. More than half of these planets are of the size of Neptune or smaller and 
the majority of them are in close-in orbits with orbital periods as small as 1.3 days 
(e.g., HD 41004 B b, \cite{Zucker04}). 

There are currently several radial velocity surveys that search for super-Earths around M stars.
Among these surveys, the Lick-Carnegie Exoplanet Survey \cite{Rivera10a,Haghighipour10,Vogt10},
the M2K Planet Search Project \cite{Clubb09,Apps10}, and the HARPS Search for Southern 
Extrasolar Planets 
\cite{Bonfils05,Udry07,Mayor09}\footnote{http://www.eso.org/sci/facilities/lasilla/instruments/harps/} 
have been successful in detecting several prominent super-Earths.

\subsection{Transit Photometry}

Similar to Doppler spectroscopy, transit photometry also shows great sensitivity to large 
planets in close-in orbits. Since in this technique, the decrease
in the light of a star is the quantity that leads to the detection of a planetary companion,
low-mass stars such as M dwarfs present more favorable targets for 
searching for transiting super-Earths.

There are currently several transit photometry surveys that use ground- and
space-based telescopes to search for Earth-like planets and super-Earths\footnote{We 
refer the reader to exoplanet.eu and exoplanets.org for more details.}. 
Among the ground-based surveys,
the MEarth project\footnote{https://www.cfa.harvard.edu/$\sim$zberta/mearth/}, a robotically 
controlled set of eight 40 cm telescopes at the F. L. Whipple Observatory on Mt. Hopkins
in Arizona, has been able to detect the first super-Earth transiting an M star (GJ 1214b, 
a 5.69 Earth-mass planet around  the M dwarf GJ 1214) \cite{Nutzman08,Irwin09a,Irwin09b,Charbonneau09}.
In space, the two telescopes CoRoT\footnote{http://smsc.cnes.fr/COROT/index.htm} and
{\it Kepler}\footnote{http://kepler.nasa.gov/} have succeeded in detecting several super-Earths
\cite{Holman10, Lissauer11} and are continuing their progress.

\begin{figure}
\begin{center}
\center{
\resizebox*{10cm}{!}{\includegraphics{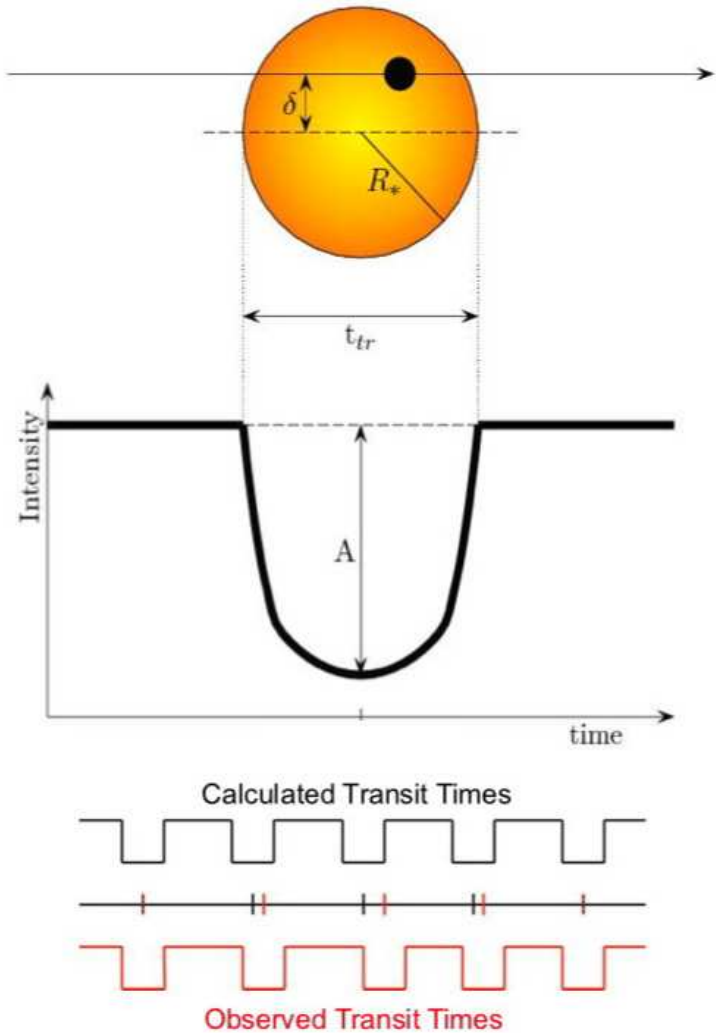}}}%
\caption{Variation of transit timing. The perturbation of an unseen planet
causes the center of transit in the observed (perturbed) system to shift from its one-planet theoretical value (black).}%
\label{TTV}
\end{center}
\end{figure}

\subsection{Transit Timing Variation Method (TTV)}

The detection of the dimming of the light of a star due to a transiting planet requires 
monitoring the star for a long time. Given the complexities in extracting information 
about a possible planetary companion from transit photometry data, and the fact
that even for planets as large and as close as hot Jupiters,
the dimming is only a few percent, it is not surprising that until recently, 
in almost all currently 
known transiting systems, the number of the detected planets was only one. 
This is, however, unlike what the simulations of planet formation suggest. 
In general, models of giant and terrestrial planet formation around
different stars indicate that planets tend to form in multiples. In other words, 
many of the currently known transiting systems may in fact harbor additional bodies \cite{Steffen10}. 
Among these (unseen) planets, those that are close to the orbit of the transiting one 
will perturb its orbit 
and cause variations in the time and duration of its transits (Figure 26).
Examples of such variations can also be observed in the transit timing of the terrestrial 
planets of our solar system \cite{Holman05}.

Measurements of the variations in the time of transit may be used to infer the existence
of the perturbing planet. As shown by \cite{Dobrovolskis96,Miralda02,Holman05,Agol05,Steffen05}, 
in a system consisting of a transiting Jupiter-like planet and a small Earth-like perturber, 
the perturbations due to the small planet create variations in the
transit timing of the hot Jupiter. The top panel of Figure 27 shows such variations 
for a Jupiter-sized 
planet in a 3-day orbit (semimajor axis of 0.047 AU) around a solar-mass star
\cite{Haghighipour08b}. The perturber in this case is 
Earth-sized and in an exterior orbit with a semimajor axis of 0.06 AU and eccentricity of 0.1. 
The amplitude of the variations in transit timing is approximately 10 seconds.

\begin{figure}
\begin{center}
\center{
\resizebox*{11cm}{!}{\includegraphics{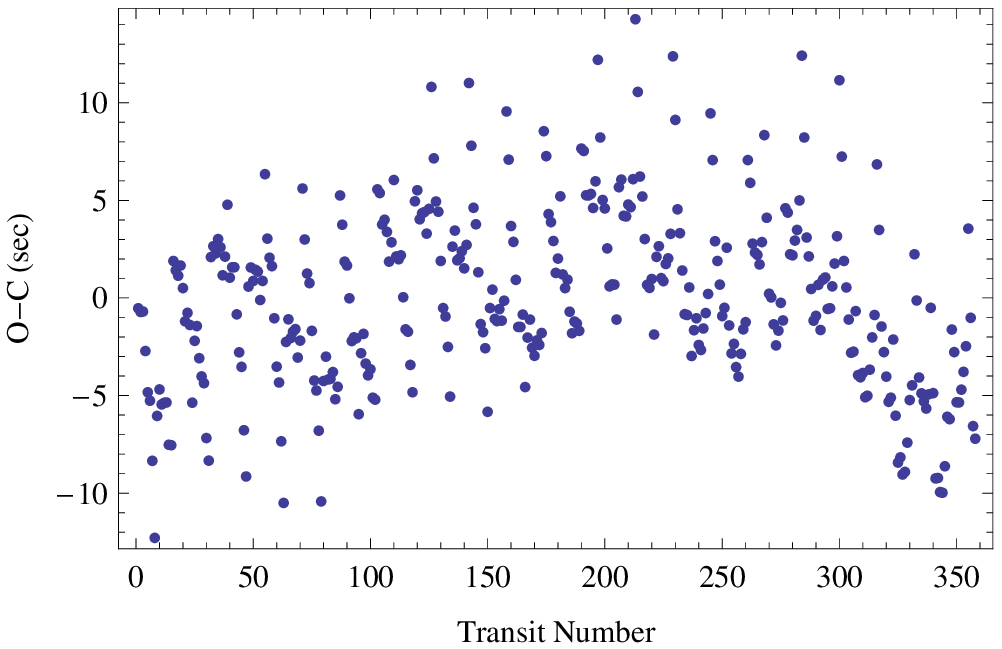}}
\vskip 10pt
\resizebox*{11cm}{!}{\includegraphics{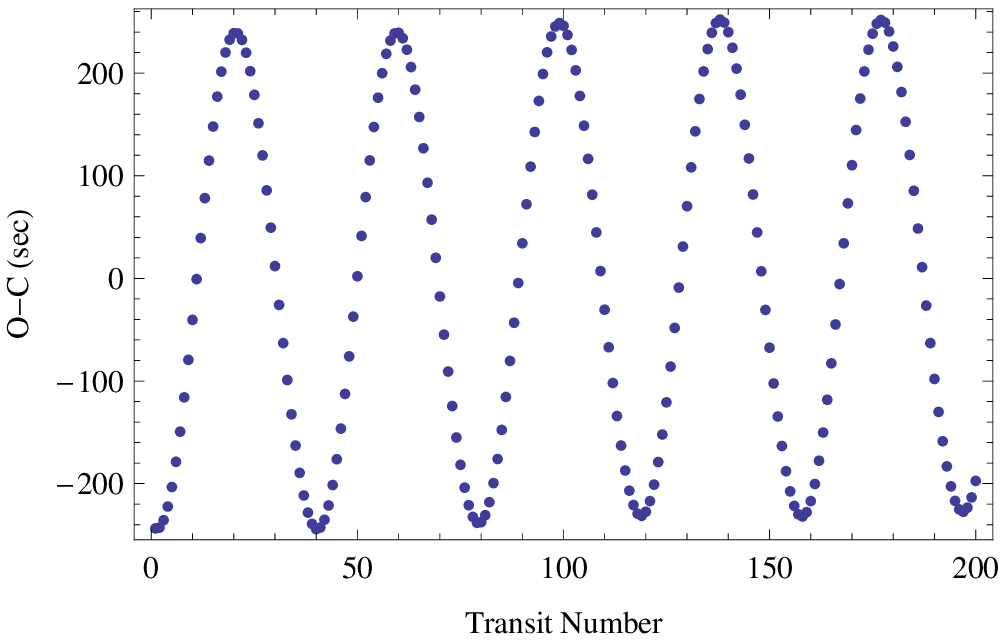}}}%
\caption{Graphs of transit timing variations for a 
transiting Jupiter-mass planet in a 3-day (0.0407 AU) circular orbit 
around a solar-mass star. The planet is perturbed by an Earth-sized object 
at 0.06 AU (top) and 0.0257 AU (bottom) where it is in an interior 2:1 MMR
with the transiting planet. The orbit of the Earth-sized planet has an 
eccentricity of 0.1. The quantity O--C represents (Observed -- Calculated).
Figure from \cite{Haghighipour08b}.}%
\label{Nader-Jason}
\end{center}
\end{figure}

Since the deviations of the orbit of a transiting planet from pure Keplerian is caused
by the gravitational force of the perturbing body, the geometrical arrangement of 
the orbits of these two planets significantly affects their mutual interactions
and the amplitude of the transit timing variations (TTVs). 
When the two planets are in or near a mean-motion resonance, 
the TTV signal is greatly enhanced. This can be seen in the bottom panel of Figure 27 where
the TTVs are shown for the same transiting planet as in the upper panel, 
but now the transiting planet and the perturber are in an interior 
2:1 mean-motion resonance (the orbital period of the perturber is 1.5 days)
\cite{Haghighipour08b}. 
A comparison between the two panels indicates that the signal in the resonant system
is more than 20 times stronger. 
This signal is, for instance, within the range of the detection sensitivity of 
{\it Kepler} space telescope.
As shown by Ford et al (2011, \cite{Ford11}),
the median transit timing uncertainty for {\it Kepler} is approximately 10 minutes
(in the system of Kepler 9 where two Saturn-sized planets are in a near
2:1 mean-motion resonance, the amplitude of the TTVs of the inner planet reaches to $\sim 50$ minutes)
and its minimum detection capability is $\sim 20$ seconds.
Recently  Payne et al (2010, \cite{Payne10}) and Haghighipour \& Kirste (2011, \cite{Kirste11}) 
studied the variation of transit timing for two-planet systems around stars with 
different masses and identified regions of the parameter-space for which
the amplitudes of TTVs can be detected by {\it Kepler} space telescope. 

Although a perturber may be able to create large disturbances in the motion of a transiting planet,
extracting information about its orbital and physical characteristics from the measurements
of the variations in the duration and intervals of transits is an extremely complicated task.
In general, observational measurements of TTVs are compared with the TTVs obtained from numerical
simulations of different planetary systems \cite{Payne10,Kirste11} and/or those calculated 
analytically \cite{Nesvorny09,Nesvorny10}. The goodness-of-fit is then used to determine
the most probable planetary system.
Unfortunately, as one can imagine, both these approaches suffer from strong degeneracy 
(different combinations of mass and
orbital elements may produce similar results). Follow-up observations, in particular with
radial velocity technique, will be necessary to confirm the detection and determine the mass 
and orbital parameters of the perturber. 

Despite these difficulties, {\it Kepler} has been able to
detected several super-Earths using the transit timing variations method \cite{Lissauer11}
(see Table 1). The study by Ford et al \cite{Ford11}, in which the transit timing of more than 
1200 potential planetary candidates in {\it Kepler}'s target list were analyzed, suggests that as
this telescope continues its successful operation,
several tens of such TTV-detected super-Earths will be identified in the near future.

\subsection{Microlensing}

While the radial velocity, transit photometry, and TTV methods are more sensitive
to detecting planets in close-in orbits, microlensing is capable of detecting 
planets in large distances. This technique has been recognized as one of the most promising 
detection methods for finding low-mass planets beyond the snow line, as well as super-Earths that 
reside further from their host stars \cite{Mann10}. Microlensing also has the unique 
capability of discovering what is known as {\it free-floating} planets. Many of these planets are 
Earth-sized or super-Earths\footnote{As the simulations of planet formation indicate, 
migrating planets, in particular the giant ones, cause the orbits of many smaller bodies
such as Earth-sized planets and super-Earths
to become unstable. These objects may be ejected from their systems 
and freely float in space.}. The Wide-Field Infrared Survey Telescope 
(WFIRST)\footnote{http://wfirst.gsfc.nasa.gov/}, NASA's future large 
space mission, will use microlensing to search for extrasolar planets. This mission 
has been recognized as the top-ranked large space mission for the next decade in the New Worlds, New Horizon 
Decadal Survey of Astronomy and 
Astrophysics\footnote{http://sites.nationalacademies.org/bpa/BPA\_049810}.  
There are currently 7 different active microlensing search projects among which 
MicroFUN (Microlensing Follow-Up Network)\footnote{http://www.astronomy.ohio-state.edu/$\sim$microfun/}, 
MOA (Microlensing Observations in Astrophysics)\footnote{http://www.phys.canterbury.ac.nz/moa/}, 
OGLE (Optical Gravitational Lensing Experiment)\footnote{http://ogle.astrouw.edu.pl/}, 
PLANET(Probing Lensing Anomalies NETwork)\footnote{http://planet.iap.fr/},
and Robonet\footnote{http://robonet.lcogt.net/} have been successful in detecting several planets.

\section*{Acknowledgments}
This manuscript is the result of discussions as well as oral and poster presentations in two NAI (NASA Astrobiology Institute) supported events: A workshop on the habitability of super-Earths held at the Aspen Center for Physics in August 2008, and a session entitled ÒBuilding Habitable PlanetsÓ held at the Goldschmidt 2009 conference. I am indebted to D. Valencia for her
invaluable contribution to this manuscript and for her constructive comments and suggestions. 
I am also thankful to E. Agol, S. Gaudi, L. Kaltenegger, S. Kenyon, as well as R. Joseph and a
second (anonymous) referee
for their critically reading of the original draft of this paper and for their valuable
suggestions that greatly improved this manuscript. Supports are also acknowledged from NASA 
Astrobiology Institute under Cooperative Agreement NNA04CC08A at the Institute for Astronomy, 
University of Hawaii, and NASA EXOB grant NNX09AN05G.

\end{document}